\documentclass[%
 reprint,
superscriptaddress,
amsmath,amssymb,
aps,
prl,
]{revtex4-1}
\usepackage{graphicx}
\usepackage{dcolumn}
\usepackage{bm}
\usepackage{multirow}

\usepackage{tabularx,array}
\makeatletter

\newcommand{\Rmnum}[1]{\expandafter\@slowromancap\romannumeral #1@}
\makeatother
\usepackage[pdftex,colorlinks,linkcolor=blue, urlcolor=blue, anchorcolor=blue, citecolor=blue]{hyperref}
\begin{document}
\maxdeadcycles=1000
\preprint{APS}
\title{Ergotropy in Quantum Batteries}

\author{Cheng-Jie Wang}
\affiliation{College of Physics and Electronic Engineering, Northwest Normal University, Lanzhou, 730070, China}
\author{Fu-Quan Dou}
\email{doufq@nwnu.edu.cn}
\affiliation{College of Physics and Electronic Engineering, Northwest Normal University, Lanzhou, 730070, China}
\affiliation{Gansu Provincial Research Center for Basic Disciplines of Quantum Physics, Lanzhou, 730000, China}

\begin{abstract}
Ergotropy--a key figure of merit for quantum battery (QB) performance--plays a crucial role. However, the dynamics and physical mechanisms governing ergotropy evolution remain open challenges. Here, we investigate the ergotropy of a general QB model and find that the charging process is accompanied by the variation and inversion of the energy level populations. In the absence of population inversion, the ergotropy is fully consistent with coherent ergotropy; in local and global population inversion, it is determined by both coherent and incoherent ergotropy. Via random sampling of quantum states and Hamiltonians, we show that coherence and the participation ratio enhance coherent ergotropy, whereas incoherent ergotropy--whether enhanced, unchanged, or suppressed--depends on diagonal entropy, the participation ratio, and energy level population ordering. We demonstrate that the ergotropy lower bound is incoherent ergotropy, the upper bound is the QB stored energy, and enhanced QB purity suppresses locked energy and boosts charging efficiency. Furthermore, we use the Tavis-Cummings (TC) and Jaynes-Cummings (JC) batteries as paradigms to validate our findings. Our work elucidates ergotropy underlying mechanisms in general QBs and establishes a rigorous framework for optimizing ergotropy and charging efficiency, paving the way for high-performance quantum energy-storage devices.
\end{abstract}
\maketitle

\emph{Introduction.}---Quantum thermodynamics focuses on the thermodynamic behavior of quantum systems, including energy conversion, work extraction, and the role of quantum resources such as quantum coherence \citep{PhysRevLett.113.140401,RevModPhys.89.041003,PhysRevA.92.022112,PhysRevLett.126.220404} and entanglement \citep{RevModPhys.81.865,PhysRevA.64.032310,PhysRevA.65.012101,PhysRevA.63.040304}. Due to the constraints imposed by the second law of thermodynamics, the stored energy in the quantum system cannot be fully extracted \citep{pnas.1411728112,PhysRevLett.115.210403}. The ergotropy \citep{Allahverdyan2004}, the maximum energy that can be extracted from a quantum system through cyclic unitary transformations, becomes one of the central issues of quantum thermodynamics \citep{PhysRevLett.130.210401}. Several appropriate characterizations of extractable work have been proposed, such as local work extraction, extended local ergotropy, and the free-energy difference \citep{PhysRevLett.133.150402,PhysRevA.107.012405}. These characterizations provide a framework to assess the impact of quantum resources, revealing that correlations induce suppression while coherence leads to enhancement in extractable work \citep{Francica2017,PhysRevLett.133.180401,PhysRevLett.125.180603}.

Quantum batteries (QBs)--one of the significant topics of quantum thermodynamics--are energy storage and extraction systems that utilize quantum mechanical principles to enhance performance or functionality \citep{PhysRevE.87.042123,RevModPhys.96.031001,QUACH20232195,Campaioli2018}. Considerable efforts has focused on constructing QBs models \citep{PhysRevB.112.104318,PhysRevA.112.L030201,PhysRevLett.134.180401,PhysRevA.110.032205,PhysRevA.110.032211,PhysRevA.106.042601,PhysRevLett.120.117702,PhysRevB.98.205423,PhysRevB.99.205437,PhysRevB.100.115142,PhysRevE.100.032107,PhysRevLett.125.236402,PhysRevA.103.052220,Dou_2020,Dou2021,PhysRevE.101.062114,PhysRevA.104.032606,PhysRevA.106.022618,PhysRevA.101.032115,PhysRevE.106.054107,PhysRevA.107.042419,PhysRevB.105.115405,PhysRevA.107.023725,PhysRevA.109.032201,PhysRevE.104.044116,PhysRevLett.132.090401,PhysRevLett.133.197001,Downing2023,PhysRevA.110.062204}, analyzing and optimizing their performance \citep{PhysRevLett.118.150601,PhysRevA.110.022433,PhysRevA.109.022226,Xu2023,PhysRevA.109.052206,PhysRevE.105.054115,PhysRevA.106.012425,PhysRevLett.127.100601,PhysRevLett.133.243602,PhysRevLett.134.220402,PhysRevE.112.024117}, and investigating the role of quantum resources \citep{PhysRevB.108.L180301,PhysRevA.107.L030201,Zhang,PhysRevB.102.245407,PhysRevA.107.032203,PhysRevLett.132.210402,PhysRevLett.131.240401,PhysRevResearch.6.023136,PhysRevResearch.2.013095,PhysRevA.106.062609}. The ergotropy is a vital figure of merit in QBs. Previous works have shown that quantum resources significantly affect the ergotropy, with coherence generally enhancing ergotropy whereas entanglement tends to suppress ergotropy  \citep{PhysRevA.111.012204,PhysRevLett.125.040601,PhysRevE.109.064103,PhysRevLett.131.060402,PhysRevE.87.042123,PhysRevE.102.052109,PhysRevLett.129.130602,PhysRevLett.134.010408,PhysRevResearch.6.023136,PhysRevA.106.062609,PhysRevA.110.052404,PhysRevLett.124.130601,PhysRevE.102.042111,9vv8-s8r1,Yang2025,PhysRevB.109.235432,PhysRevB.104.245418,PhysRevLett.122.047702}. Further theoretical and experimental investigations have demonstrated that ergotropy consists of both coherent and incoherent contributions and that the coherent ergotropy can be promoted by coherence \citep{PhysRevLett.125.180603,PhysRevLett.133.180401,PhysRevLett.129.130602}. The bounds of ergotropy are closely related to the battery capacity, and are determined by von Neumann's trace inequality \citep{PhysRevLett.131.030402,PhysRevA.109.042424,Caravelli2021}. Another performance metric is the charging efficiency, defined as the ratio between the ergotropy and the stored energy \citep{PhysRevB.109.235432}. The higher-energy chargers and larger ensembles of battery units can improve this efficiency \citep{PhysRevB.100.115142,RevModPhys.96.031001,PhysRevA.109.052206,PhysRevA.112.022217,PhysRevA.106.062609,vqnk-kzqg}. Nevertheless, these results are typically model dependent. Within a universal framework of QBs, the dynamical evolution of ergotropy (including coherent and incoherent ergotropy), their bounds, and the underlying physical mechanisms remain challenges. Therefore, it is of fundamental importance to address thoroughly the following issues: (i) How do coherent and incoherent ergotropy govern the ergotropy? (ii) How are coherent and incoherent ergotropy influenced by quantum resources, and how can one characterize the bounds of the ergotropy, in particular those of the coherent contribution? (iii) How do factors affect the charging efficiency, and how can this efficiency be further improved?

In this Letter, we analyze the dynamics and physical mechanisms of ergotropy in model-independent QBs, based on the fact that ergotropy consists of coherent and incoherent components. By means of random or weighted random sampling of quantum states and their associated Hamiltonians \citep{PhysRevLett.134.010406,SupplementalMaterial}, we investigate the relationship between the coherent and incoherent contributions on ergotropy and energy-level populations, as well as the impact of quantum resources on both ergotropy components, and the bounds of the ergotropy. We find that both coherence and participation ratio enhance the coherent ergotropy, while diagonal entropy (entanglement) and participation ratio can enhance, remain or suppress the incoherent ergotropy. Additionally, we demonstrate the role of purity in decreasing the locked energy and improving the charging efficiency. We illustrate our major results in the Tavis-Cummings (TC) and Jaynes-Cummings (JC) batteries. These results offer key insights into the design of high-performance QBs, optimizing energy extraction and charging efficiency through quantum resources.

\emph{Preliminaries.}\label{section1}---We consider the general QB composed by a battery and a charger. The charging process is governed by the following Hamiltonian
\begin{equation}\label{H}
H\left(t\right)=H_B+H_C+\lambda\left(t\right)\mathcal{V},
\end{equation}
where $H_B$ and $H_C$ are the Hamiltonians of the battery and charger, respectively. The charging operator $\mathcal{V}$ contains all terms that control energy injection, such as charger-battery interactions and some external driven fields. The time-dependent parameter $\lambda\left(t\right)$ describes the charging period, and we assume it follows a step function that equals to 1 for $t\in\left[0,T\right]$ and zero elsewhere. At time $t=0$, the charger-battery system is prepared in a product state $\rho_{BC}\left(0\right)$ with the battery being the ground state $|\varepsilon_1\rangle$ of $H_B$. In a finite time interval $\left[0,T\right]$, the evolving state is given by $\rho_{BC}\left(t\right)=U\left(t\right)\rho_{BC}\left(0\right)U^\dagger\left(t\right)$ with $U\left(t\right)=\mathcal{T}\exp[-i\int_0^t H\left(\tau\right) d\tau/\hbar]$, where $\mathcal{T}$ is a time-ordering operator. The stored energy in QB at time $t$ is defined by
\begin{equation}\label{EB}
E\left(\rho_B\right)=\mathrm{Tr}\left[H_B\rho_{B}\left(t\right)\right]-\mathrm{Tr}\left[H_B\rho_B\left(0\right)\right],
\end{equation}
where $\rho_B\left(t\right)=\mathrm{Tr}_C\left[\rho_{BC}\left(t\right)\right]$ is the reduced density matrix of the battery. The ergotropy is obtained by the difference between the stored energy $E(\rho_B)$ and locked energy $E(\tilde{\rho}_B)$
\begin{equation}\label{ergotropy}
\mathcal{E}\left(\rho_B\right)=\mathrm{Tr}\left[H_B\rho_{B}\right]-\mathrm{Tr}[H_{B}\widetilde{\rho}_{B}],
\end{equation}
where $\widetilde{\rho}_B=\sum_{n=1}^d r_n|\varepsilon_n\rangle\langle \varepsilon_n|$ is the passive state, $r_n$ are the eigenvalues of $\rho_B$ ordered as $r_1\geq r_2\geq\cdots\geq r_d$, and $|\varepsilon_n\rangle$ are the eigenstates of $H_B$ with the corresponding eigenenergies $\varepsilon_n$, which satisfy $\varepsilon_1\leq \varepsilon_2\leq\cdots\leq \varepsilon_d$. Here, $d$ denotes the dimension of the battery system. The incoherent ergotropy quantifies the maximum work unitarily extractable from $\rho_B$ without altering its coherence and is written as
\begin{equation}\label{incoer}
\mathcal{E}_i\left(\rho_B\right)=\mathrm{Tr}\left\{H_B\delta_B\right\}-
\mathrm{Tr}\left\{H_B\widetilde{\delta}_B\right\},
\end{equation}
where $\delta_{B}=\sum_{n=1}^dp_n|\varepsilon_n\rangle\langle\varepsilon_n|$ is the dephased state of $\rho_B$ where $p_n$ denotes the population of $n$th energy level, $\widetilde{\delta}_{B}=\sum_{n=1}^d\tilde{p}_n|\varepsilon_n\rangle\langle \varepsilon_n|$ is the passive state of $\delta_B$ with $\tilde{p}_{n}\geq\tilde{p}_{n+1}$. Having defined the incoherent component of $\mathcal{E}$, the coherent contribution to ergotropy can be readily given by the difference
\begin{equation}\label{coer}
\mathcal{E}_c\left(\rho_B\right) = \mathcal{E}\left(\rho_B\right)-\mathcal{E}_i\left(\rho_B\right)=\mathrm{Tr}\left[\tilde{\delta}_BH_B\right]-\mathrm{Tr}\left[\tilde{\rho}_BH_B\right].
\end{equation}
The charging efficiency as another useful quantifier of QBs performance is defined as
\begin{equation}\label{ratio}
\mathcal{R}\left(\rho_B\right)=\frac{\mathcal{E}\left(\rho_B\right)}{E\left(\rho_B\right)}.
\end{equation}

Quantum coherence and entanglement, the fundamental quantum resources, play an important role in work extraction from quantum systems. The coherence of state $\rho_B$ is quantified by the quantum relative entropy of $\rho_B$ \citep{PhysRevLett.113.140401}:
\begin{equation}\label{coherence}
\mathcal{C}\left(\rho_B\right)=S_{\text{diag}}\left(\rho_B\right)-S\left(\rho_B\right),
\end{equation}
where $S\left(\rho\right)=-\mathrm{Tr}\left[\rho \log\left(\rho\right)\right]$ is the von Neumann entropy, and  $S_{diag}\left(\rho_B\right)=-\sum_{n=1}^dp_n\log{p_n}$ is the diagonal entropy. For any pure bipartite state $\rho_{BC}$, the von Neumann entropy of the battery's reduced density matrix $S\left(\rho_B\right)$ characterizes the charger-battery entanglement. When the coherence vanishes, the diagonal entropy equals to the von Neumann entropy, thereby characterizing the entanglement between the battery and charger. To further explore the connection between ergotropy and quantum resources, the participation ratio
\begin{equation}\label{PR}
\mathrm{PR}\left(\rho_B\right)=\frac{1}{\sum_{n=1}^dp_n^2},
\end{equation}
a measure of the localization of quantum states, and purity
\begin{equation}\label{pur}
\mathcal{P}\left(\rho_B\right) = \mathrm{Tr}\left[\rho_B^2\right],
\end{equation}
a natural indicator of the mixedness of the battery state, are introduced. These quantities provide a comprehensive framework to assess the interplay between population distribution, quantum resources, and ergotropy in QBs.

\begin{figure}[htbp]
\centering
\includegraphics[width=0.485\textwidth]{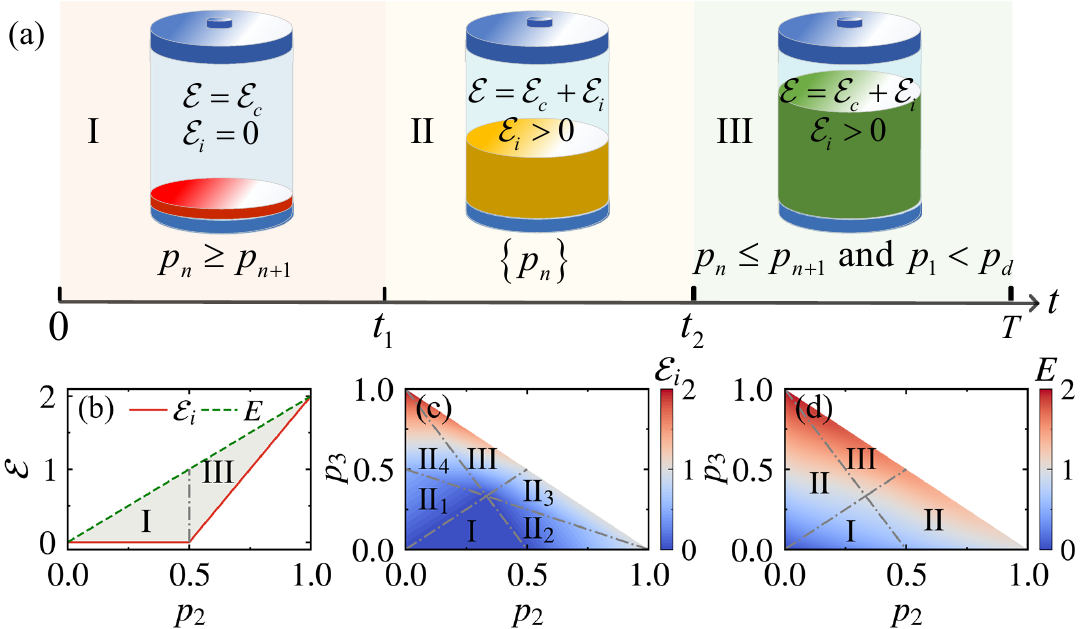}
\caption{(a) Schematic diagram of the charging process, which consists of three stages: Stage $\mathrm{\Rmnum{1}}$, no population inversion; Stage $\mathrm{\Rmnum{2}}$, local population inversion; Stage $\mathrm{\Rmnum{3}}$, global population inversion. The other parameters satisfy $0\leq t_1\leq t_2\leq T$ and $T>0$. (b) The ergotropy as a function of the excited energy level population $p_2$ for two-level QBs. (c),(d) The incoherent ergotropy and the stored energy as a function of the population of $2$th and $3$th energy level, $p_2,p_3$ for three-level QBs. In (c), the region of local population inversion contains different population ordering: In $\mathrm{\Rmnum{2}}_1$, the population ordering is $ p_1 \geq p_3 > p_2 $; in $\mathrm{\Rmnum{2}}_2$, $p_2 > p_1 \geq p_3$; in $\mathrm{\Rmnum{2}}_3$, $p_2 > p_3 \geq p_1$ or $ p_2 \geq p_3 > p_1 $; and in $\mathrm{\Rmnum{2}}_4$, $ p_3 \geq p_1 > p_2 $ or $p_3 > p_1 \geq p_2$. Here, the Hamiltonian $H_B$ has equally spaced energy levels for two-level and three-level QBs.}
\label{fig1}
\end{figure}

\emph{Proposition~}1.\label{section2}---In the absence of population inversion, the ergotropy coincides with the coherent ergotropy--with the incoherent ergotropy being zero--i.e., $\mathcal{E}=\mathcal{E}_c$, $\mathcal{E}_i=0$. In the presence of population inversion, however, the ergotropy is governed by both coherent and incoherent contributions, such that $\mathcal{E}=\mathcal{E}_c+\mathcal{E}_i$.

Population evolution plays a crucial role in the charging process, which is generally accompanied by population changes and inversion. As sketched in Fig. \ref{fig1}(a), the charging process can usually be divided into three distinct cases or stages: ($\mathrm{\Rmnum{1}}$) The no population inversion, where the population is ordered in a descending sort, i.e., $\forall\ 1\leq n< m\leq d$, $p_n\geq p_m$; ($\mathrm{\Rmnum{2}}$) The local population inversion, characterized by the population inversion involving some energy levels, i.e., $\exists\ 1\leq n<m \leq d$, $p_n<p_m$; ($\mathrm{\Rmnum{3}}$) The global population inversion, where the population is ordered in an ascending sort, i.e., $\forall\ 1\leq n<m\leq d$, $p_n\leq p_m$ and $p_1<p_d$. For two-level QBs, the process only involves two stages, namely $\mathrm{\Rmnum{1}}$ and $\mathrm{\Rmnum{3}}$.

In the stage with no population inversion, the dephased state coincides with the passive state, $\delta_B = \tilde{\delta}_B$, which indicates that the incoherent ergotropy vanishes $\mathcal{E}_i=0$, and the ergotropy reduces to the coherent contribution, namely $\mathcal{E}=\mathcal{E}_c$. Conversely, in the stage with local or global population inversion, the dephased state has higher energy than the passive state, i.e., $\mathrm{Tr}[\delta_BH_B]>\mathrm{Tr}\left[\tilde{\delta}_BH_B\right]$, resulting in a nonzero incoherent ergotropy and a total ergotropy $\mathcal{E}=\mathcal{E}_c+\mathcal{E}_i$.

In Figs. \ref{fig1}(b)-\ref{fig1}(d), we illustrate the above results in two-level and three-level QBs (it remains valid in the high-dimensional case). 
 Here, we normalize the Hamiltonian of battery as $1/\left(\varepsilon_d-\varepsilon_1\right)\left[2H_B-\left(\varepsilon_d+\varepsilon_1\right)\mathbb{I}\right]\rightarrow H_B$ \citep{PhysRevA.109.042207}. In the stage with no population inversion corresponding to the region $\mathrm{\Rmnum{1}}$, the coherent part governs the ergotropy. In the local population inversion depicted in the region $\mathrm{\Rmnum{2}}$, both incoherent and coherent ergotropy jointly control the ergotropy, whereas in the global population inversion corresponding to the region $\mathrm{\Rmnum{3}}$, the incoherent part gradually becomes dominant as the populations of higher-energy levels increase. Until the population of the highest energy level  $p_2$ in Fig. \ref{fig1}(b) or $p_3$ in Figs. \ref{fig1}(c) and \ref{fig1}(d) reaches $1$, the ergotropy is consistent with the incoherent ergotropy. Furthermore, we obtain an important corollary: the lower bound of ergotropy is the incoherent ergotropy, and the upper bound is the stored energy of battery for any given population distribution.

This proposition reveals the dynamical feature of ergotropy from perspective of coherent and incoherent contributions. A natural question arises: what is the physical mechanism underlying the coherent and incoherent ergotropy, i.e., which quantum resources or factors influence the coherent and incoherent parts?

\emph{Proposition }2.\label{section3}---(i) The coherent ergotropy-enhanced by both coherence and participation ratio-exhibits well-defined upper and lower bounds, which correspond to the maximally delocalized state and pure state limits, respectively. (ii) The incoherent ergotropy--being enhanced, maintained, or suppressed--depends on the diagonal entropy or participation ratio, and the ordering of level populations. For each ordering, there exist upper and lower bounds of the incoherent ergotropy, as well as a series of special points of the diagonal entropy $\log n$: extremum and corner (critical) points, across which the boundary behavior of the incoherent ergotropy may change.

To explore the roles of coherent and incoherent factors on the coherent ergotropy, we characterize the coherent and incoherent factors as coherence and participation ratio, respectively. We first investigate the effect of the coherence and participation ratio on the coherent ergotropy through a large sample of randomly generated states on a single randomly generated $H_B$. Here, the states are drawn from the Hilbert-Schmidt measure, and $H_B$ is sampled from the Gaussian unitary ensemble [see Supplemental Material (SM) \citep{SupplementalMaterial} for details]. The results are illustrated in Figs. \ref{fig2}(a) and \ref{fig2}(b), which reveal a positive dependence of coherent ergotropy on the coherence and participation ratio. The coherent ergotropy, $\mathcal{E}_c = \mathrm{Tr}\left[\tilde{\delta}_BH_B\right]-\mathrm{Tr}\left[\tilde{\rho}_B H_B\right]$, can provide support for these results. For a given coherence, a larger participation ratio indicates a higher locked energy $\mathrm{Tr}\left[\tilde{\delta}_BH_B\right]$ of the dephased state $\delta_B$ and thereby potentially enhancing the coherent ergotropy \citep{SupplementalMaterial}; for a given participation ratio, a greater coherence implies higher purity of $\rho_B$ and the lower locked energy $\mathrm{Tr}\left[\tilde{\rho}_B H_B\right]$ (see also Proposition $3$), which likewise contributes to increased coherent ergotropy. Due to that the participation ratio of the pure state $\rho_B^{\text{pur}}$ is minimal and that of the completely delocalized state $\rho_B^{\text{deloc}}$ \citep{Karamlou2022} is maximal for any given coherence, we can obtain bounds for the coherent ergotropy $\mathcal{E}_c$ \citep{SupplementalMaterial}:
\begin{equation}\label{coerbound}
\mathcal{E}_c\left(\rho_B^{\text{pur}}\right)\leq \mathcal{E}_c\left(\rho_B\right)\leq \mathcal{E}_c\left(\rho_B^{\text{deloc}}\right),
\end{equation}
where
$\mathcal{E}_c\left(\rho_B^{\text{pur}}\right)=\frac{1}{\alpha} \left[\mathcal{C}\left(\rho_B^\text{pur}\right)+\sum_{n=1}^d\tilde{p}_n\log\chi_n-\alpha\varepsilon_1\right]$, $\mathcal{E}_c\left(\rho_B^{\text{deloc}}\right)=\frac{1}{\alpha}\Big[\mathcal{C}\left(\rho_B^\text{deloc}\right)+\alpha\mathrm{Tr}\left[H_B\rho_B^\text{deloc}\right]-\log {d}-\sum_{n=1}^dr_n\log{q_n}\Big]$,
and $\mathcal{C}(\rho_B^{\text{pur}}) = \mathcal{C}(\rho_B^{\text{deloc}}) = \mathcal{C}(\rho_B)$. The parameters $r_n=q_ne^{-\alpha \varepsilon_n}$ and $\tilde{p}_n=\chi_ne^{-\alpha\varepsilon_n}$ are the eigenvalues of $\rho_B^{\text{pur}}$ and the dephased state corresponding to $\rho_B^{\text{deloc}}$ where $\alpha,q_n,\chi_n$ are auxiliary parameters. We display the bounds of coherent ergotropy for two- and three-levels QBs in Figs. \ref{fig2}(a) and \ref{fig2}(b), respectively. In the two-level QBs, they are represented as a distinct line, whereas in the three-level QBs, they appear as a band. In comparison to the bounds reported in Ref. \citep{PhysRevLett.125.180603}, our results are tighter.

\begin{figure}[htbp]
\centering
\includegraphics[width=0.485\textwidth]{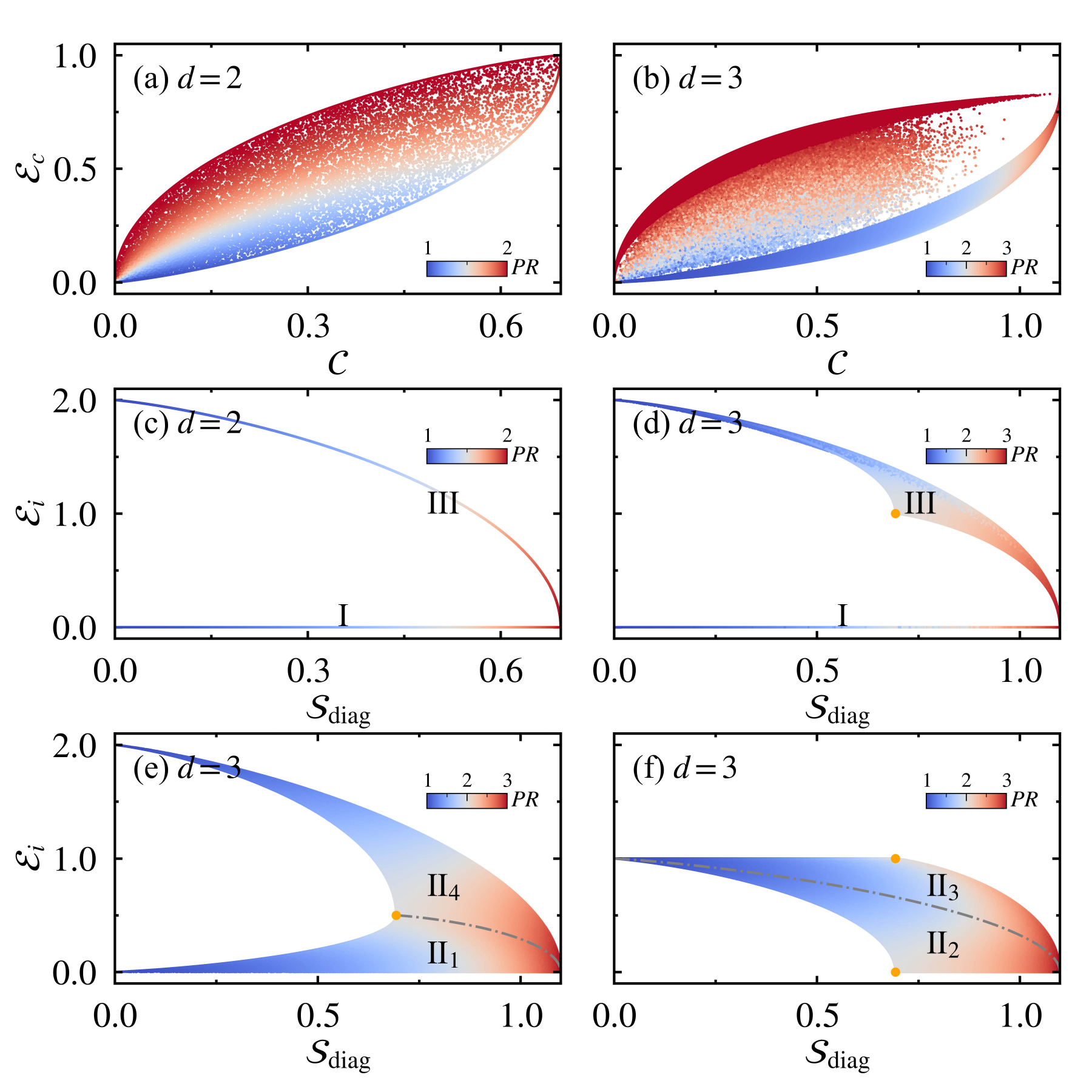}
\caption{(a),(b) The coherent ergotropy $\mathcal{E}_c$ as a function of the coherence $\mathcal{C}$ and participation ratio $PR$ for randomly sampled states $\rho_B$, Hamiltonian $H_B$ for two-level and three-level QBs. Here, the Hamiltonian $H_B$ is normalized. (c)-(f) For the cases of the no, local and global population inversion the incoherent ergotropy as a function of the diagonal entropy and participation ratio for randomly sampled states $\rho_B$ and Hamiltonian $H_B$. The symbols $\mathrm{\Rmnum{1}}$, $\mathrm{\Rmnum{2}}_{1}$, $\mathrm{\Rmnum{2}}_{2}$, $\mathrm{\Rmnum{2}}_{3}$, $\mathrm{\Rmnum{2}}_{4}$ and $\mathrm{\Rmnum{3}}$ denote the same configurations as in Fig. \ref{fig1}(c). The orange dots in (e) and (f) correspond to the critical value of diagonal entropy, $\log 2$. We focus on Hamiltonians with equally spaced and normalized energy levels. Panels (a) and (b) contain $10^{5}$ randomly sampled states, while panels (c)-(f) contain $10^{7}$ states.
}
\label{fig2}
\end{figure}

We now turn to the incoherent ergotropy. Since the incoherent ergotropy is governed solely by incoherent factors, we investigate the effects of diagonal entropy and participation ratio on the incoherent ergotropy through the weighted random sampling of $\rho_B$ and $H_B$ \citep{SupplementalMaterial}, as shown in Figs. \ref{fig2}(c)-\ref{fig2}(f) for two- and three-levels QBs. According to Proposition $1$, the incoherent ergotropy remains zero in the absence of population inversion [see also Figs. \ref{fig1}(b) and \ref{fig1}(c)]. We mainly focus on the cases of the global and local population inversion. For the two-level QBs, the incoherent ergotropy is negatively related to the diagonal entropy or participation ratio in the case of global population inversion, as depicted in Fig. \ref{fig2}(c). For the three-level QBs, the relationship between the incoherent ergotropy and the diagonal entropy or participation ratio is no longer monotonic; it may increase, remain, or suppress under different conditions (for example, when the variation of energy-level population satisfies $\Delta p_1>\max \{-1/2\Delta p_2,-\Delta p_2\log\frac{p_3}{p_2}/\log\frac{p_3}{p_2}\}$ or $\Delta p_1<\min \{-1/2\Delta p_2,-\Delta p_2\log\frac{p_3}{p_2}/\log\frac{p_3}{p_2}\}$, the diagonal entropy suppresses the incoherent ergotropy; a detailed proof can be found Ref. \citep{SupplementalMaterial}). Nevertheless, its bounds remain negatively related with the diagonal entropy (participation ratio). Therefore, to achieve larger incoherent ergotropy, smaller diagonal entropy or participation ratio is generally preferred in the case of global population inversion. For the case of the local population inversion, the relation is similar to that for global population inversion. For each ordering, the incoherent ergotropy has upper and lower bounds, as well as extremum and critical points of the diagonal entropy. On the one hand, when $\mathcal{S}_{\text{diag}}=0$, $\rho_B$ is localized in an arbitrary eigenstate $|\varepsilon_n\rangle$, i.e., $\mathcal{E}_i=\varepsilon_n-\varepsilon_1$; when $\mathcal{S}_{\text{diag}}=\log d$, the population is equally distributed across all energy levels in $H_B$, i.e., $\mathcal{E}_i=0$. On the other hand, a critical point may occur where the diagonal entropy is $\log n$ (with $n=2,3,...,d-1$), across which the boundary behavior may change. For example, in $\mathrm{II}_2$ region of Fig. \ref{fig2}(e), before the critical point, the upper bound of incoherent ergotropy increases with the diagonal entropy, while after the critical point, it is suppressed. We provide the detailed proof of the conditions under which the diagonal entropy may enhance, remain, or suppress the incoherent ergotropy in SM \citep{SupplementalMaterial}.

This proposition reveals how quantum resources can be exploited to enhance both coherent and incoherent ergotropy. Not only ergotropy but also charging efficiency should be considered to fully assess QB performance. Given that the stored energy can be fully extracted in the pure state, a natural question arises as to whether increasing purity enhances the charging efficiency.

\emph{Proposition 3.}--\label{section4}Purity generally suppresses the locked energy $\mathrm{Tr}\left[\tilde{\rho}_BH_B\right]$ while it enhances the charging efficiency $\mathcal{R}$ (in the presence of population inversion) during the charging process.

For energy storage devices, the charging efficiency represents another crucial performance metric.~To realize high-performance QBs, we now explore how purity can be utilized to reduce the locked energy and enhance the charging efficiency. The purity of $\rho_B$ remains invariant under unitary evolution: $\mathcal{P}(\rho_B) = \mathrm{Tr}[\rho_B^2] = \mathrm{Tr}[U\rho_B U^\dagger U\rho_B U^\dagger] = \mathrm{Tr}[\tilde{\rho}_B^2] = \sum_{n=1}^d r_n^2$. For the passive state $\tilde{\rho}_B$, the purity and participation ratio satisfy $\mathcal{P}(\tilde{\rho}_B) = 1/\mathrm{PR}(\tilde{\rho}_B)$. This implies that an increase in purity of $\rho_B$ enhances the localization of $\tilde{\rho}_B$. Due to $r_1 \geq r_2 \geq \dots \geq r_d$, stronger localization increases the population of $\tilde{\rho}_B$ in lower energy levels, thereby reducing locked energy $E(\tilde{\rho}_B)$. Since the charging efficiency is given by $\mathcal{R}(\rho_B) = 1 - E(\tilde{\rho}_B)/E(\rho_B)$, a decrease in the locked energy may enhance the charging efficiency. Figure \ref{fig3} illustrates the dependence of both the locked energy and the charging efficiency (under population inversion) on  purity of two- and three-levels QBs, based on a large sample of randomly generated states on a single randomly generated Hamiltonian \citep{SupplementalMaterial}. Purity typically suppresses the locked energy, particularly in the two-level QBs where the locked energy is negatively related to the purity [see Figs. \ref{fig3}(a) and \ref{fig3}(b)]. In addition, purity generally enhances the charging efficiency, as observed for each population ordering in the tow- and three-level QBs [see Figs. \ref{fig3}(c) and \ref{fig3}(d)]. Notably, when the battery is in a pure state, i.e., $\mathcal{P}(\rho_B) = 1$, the charging efficiency reaches its maximum value $1$, indicating that the stored energy in the battery can be fully extracted. Consequently, increasing the purity can enable higher charging efficiency.
\begin{figure}[tbp]
\centering
\includegraphics[width=0.485\textwidth]{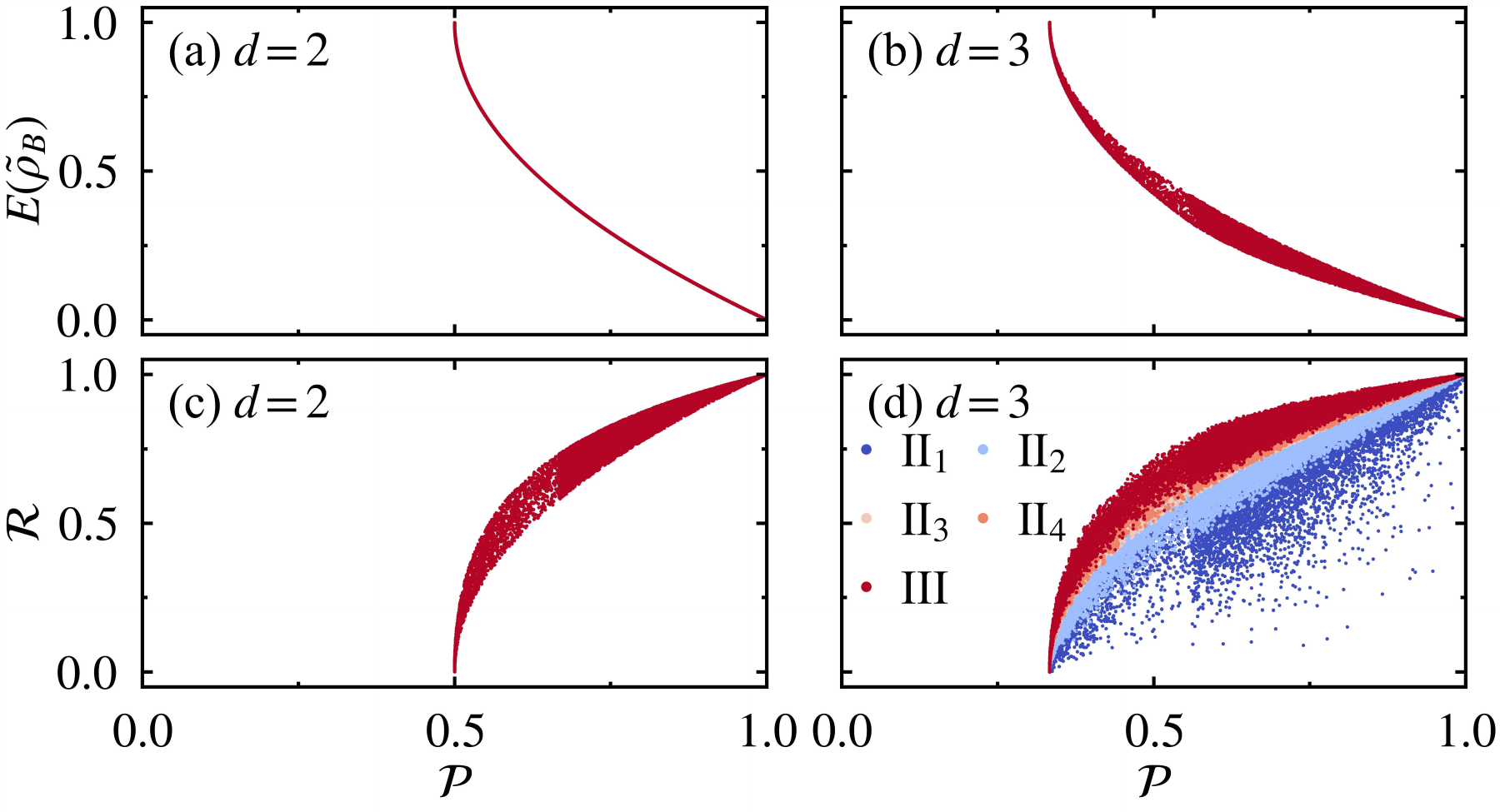}
\caption{The dependence of both (a),(b) the locked energy $E(\tilde{\rho}_B)$ and (c),(d) the charging efficiency $\mathcal{R}$ on purity $\mathcal{P}$ for two-level and three-level QBs. In (d), the symbols $\mathrm{\Rmnum{2}}_{1}$, $\mathrm{\Rmnum{2}}_{2}$, $\mathrm{\Rmnum{2}}_{3}$, $\mathrm{\Rmnum{2}}_{4}$ and $\mathrm{\Rmnum{3}}$ denote the same configurations as in Fig. \ref{fig1}(c). Here, the Hamiltonian $H_B$ is normalized. Each panel contains $10^5$ randomly sampled states.
}
\label{fig3}
\end{figure}

\emph{Example: JC battery and TC battery.}\label{section5}---The TC battery consists of $N_B$ two-level atoms and a single-mode cavity. The charger-battery system can be described by the following Hamiltonian (we set $\hbar=1$)
\begin{equation}\label{HTC}
H=\omega_c a^\dagger a+\omega_b J_z+g(J_+ a+J_-a^\dagger),
\end{equation}
where $H_C=\omega_c a^\dagger a,\ H_B=\omega_b J_z$ and $\mathcal{V}=g(J_+a+J_-a^\dagger)$ are the Hamiltonian of charger, battery and their interaction, respectively. The $a$ ($a^\dagger$) annihilates (creates) a photon in cavity with frequency $\omega_c$ and $g$ characterizes the strength of coupling. Here, $J_\alpha =1/2 \sum_{n=1}^{N_B}\sigma_n^\alpha\ (\alpha=x,y,z)$ is a collective spin operator and $J_\pm=J_x\pm iJ_y$ are spin ladder operators. We consider the resonant scenario, i.e., $\omega_c=\omega_b=\omega$. When the number of two-level atom is one, the Hamiltonian (\ref{HTC}) can describe the JC battery. The initial state of system is the product state $\mid\psi\left(0\right)\rangle = \mid 0 \rangle_B\otimes\mid\alpha\rangle_C$, where the battery is in the ground state of $H_B$, i.e., $\mid0\rangle_B=\mid\downarrow,\dots,\downarrow\rangle$ and the charger is in a coherent state $\mid\alpha\rangle_C$ with $N_c$ photons.
\begin{figure}[htbp]
\centering
\includegraphics[width=0.485\textwidth]{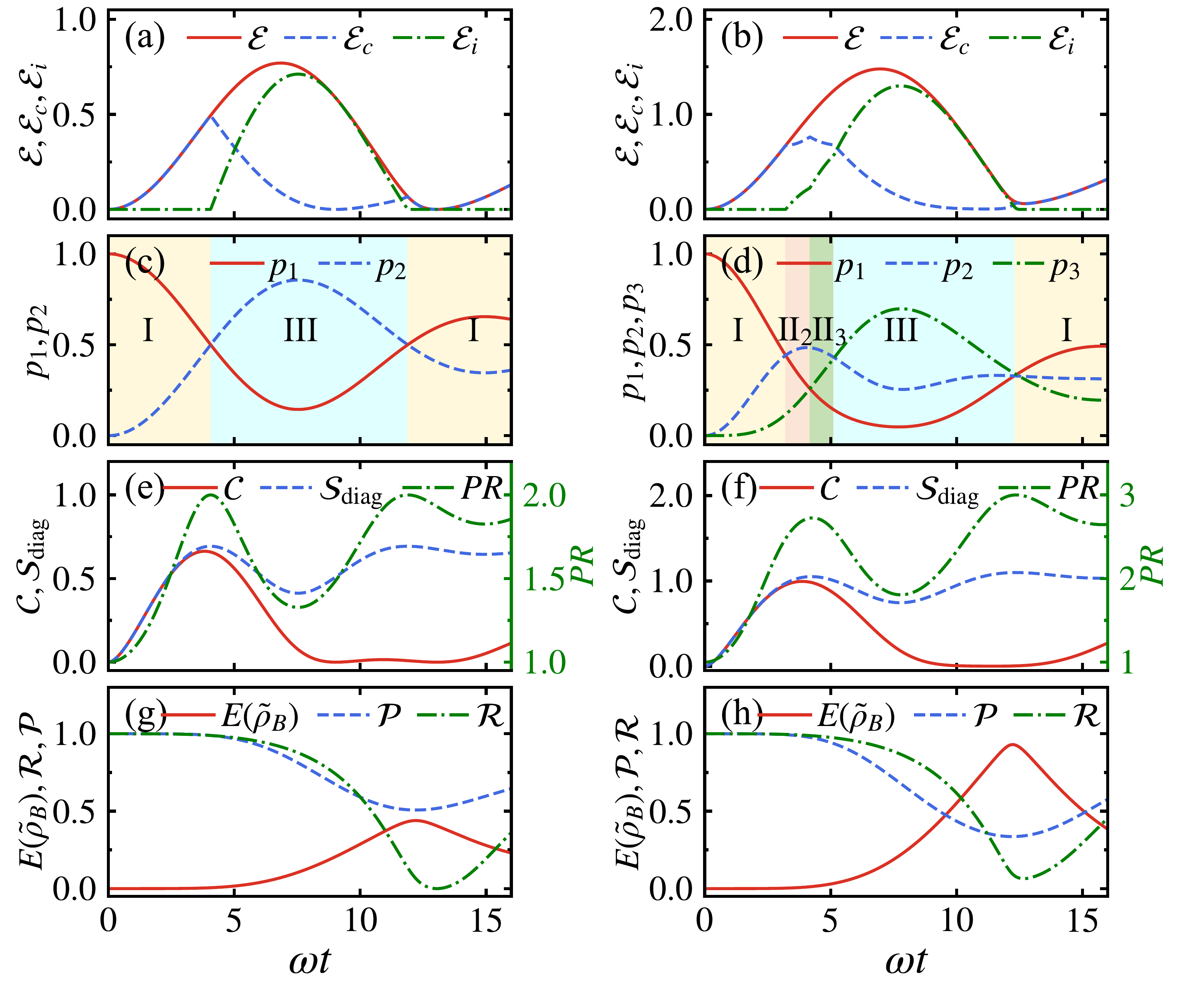}
\caption{Time evolution of (a),(b) ergotropy, coherent ergotropy, incoherent ergotropy, (c),(d) energy level populations, (e),(f) coherence, diagonal entropy, participation ratio, (g),(h) locked energy, purity and charging efficiency. (a), (c), (e) and (g) correspond to the case of JC QBs, and (b), (d), (f) and (h) correspond to the case of TC QBs containing two atoms ($N_B=2$), which is analogous to the three-level QBs. The symbols $\mathrm{\Rmnum{1}}$, $\mathrm{\Rmnum{2}_2}$, $\mathrm{\Rmnum{2}}_3$, $\mathrm{\Rmnum{3}}$ denote the same setting as in Fig. \ref{fig1}(c). Other parameters are set to $\omega=1$, $g=0.1$ and $N_c=4$.}
\label{fig4}
\end{figure}

Time evolution of ergotropy, coherent ergotropy and incoherent ergotropy are shown in Figs.~\ref{fig4}(a) and \ref{fig4}(b) for JC battery and TC battery, respectively. Combining the energy level populations of JC battery and TC battery [see Figs. \ref{fig4}(c) and \ref{fig4}(d)], it supports the Proposition $1$ that $\mathcal{E}=\mathcal{E}_c,\ \mathcal{E}_i=0$ without the population inversion and $\mathcal{E} = \mathcal{E}_c+\mathcal{E}_i$ with the population inversion. Further, we demonstrate the influence of quantum resource (quantum coherence, diagonal entropy and participation ratio) on both the coherent ergotropy and incoherent ergotropy in Figs. \ref{fig4}(a), \ref{fig4}(b), \ref{fig4}(e) and \ref{fig4}(f). The coherent ergotropy increases with the coherence and participation ratio. The incoherent ergotropy can be enhanced and suppressed by the diagonal entropy or participation ratio. In the case of global population inversion (the $\mathrm{\Rmnum{3}}$ region), for JC and TC batteries the incoherent ergotropy is negatively related to the diagonal entropy or participation ratio and the maximal value of incoherent ergotropy corresponds to the minimal value of the diagonal entropy or participation ratio. In case of local population inversion (the $\mathrm{\Rmnum{2}}_2$ region), the diagonal entropy enhances the incoherent ergotropy. In addition, the dependence of both the locked energy and the charging efficiency on the purity is illustrated in Figs. \ref{fig4}(g) and \ref{fig4}(h). It is evident that the purity suppresses the locked energy and enhances the charging efficiency. For the different initial states and the open QBs, the above results still hold \citep{SupplementalMaterial}.

\emph{Conclusions.}---We have demonstrated the coherent and incoherent contributions on ergotropy and the impact of quantum resource on them for general QBs based on the energy level populations. During the charging process, the ergotropy coincides with the coherent ergotropy in the absence of population inversion and is determined by both the coherent and incoherent ergotropy in the presence of population inversion. For any given population, the lower bound of ergotropy is the incoherent ergotropy, and its upper bound is the stored energy. The coherence and participation ratio jointly can enhance the coherent ergotropy, whose upper and lower bounds are attained for the maximally delocalized state and the pure state, respectively. The diagonal entropy and participation ratio can enhance, remain or suppress the incoherent ergotropy under different conditions. In the cases of global population inversion, the diagonal entropy (participation ratio) generally suppresses the incoherent ergotropy and the minimal value of diagonal entropy (participation ratio) corresponds to the maximal value of the incoherent ergotropy. We have also illustrated that the purity generally can suppress the locked energy and enhance the charging efficiency. Our results elucidate a clear ergotropy mechanism in general QBs and provide a valuable guidance for optimal ergotropy and charging efficiency of QBs, advancing high-performance quantum energy-storage devices.

\emph{Acknowledgments.}---The work is supported by the National Natural Science Foundation of China (Grant No. 12475026) and the Natural Science Foundation of Gansu Province (No. 25JRRA799).
\bibliography{reference}
\onecolumngrid
\newpage

\begin{center}
    \Large\textbf{Supplemental Material for ``Ergotropy in Quantum Batteries''}
\end{center}

\begin{center}
Cheng-Jie Wang\textsuperscript{1}, and Fu-Quan Dou\textsuperscript{1,2,*}\\
\vspace{5pt}
\small
\textsuperscript{1}\emph{College of Physics and Electronic Engineering, Northwest Normal University, Lanzhou, 730070, China}\\
\textsuperscript{2}\emph{Gansu Provincial Research Center for Basic Disciplines of Quantum Physics, Lanzhou, 730000, China}

\end{center}
\vspace{25pt}
\setcounter{section}{0}
\renewcommand{\thesection}{\Roman{section}}
\setcounter{figure}{0}
\renewcommand{\thefigure}{S\arabic{figure}}
\setcounter{equation}{0}
\renewcommand{\theequation}{S\arabic{equation}}
\setcounter{table}{0}
\renewcommand{\thetable}{S\arabic{table}}

\noindent
This Supplemental Material provides a comprehensive description of the numerical methods employed and detailed derivations of all the analytical results in the main text.  Section I introduces all the details of our random sampling methods. Then we analyze the relation between the locked energy $E(\tilde{\delta}_B)$ and participation ratio $PR(\tilde{\delta}_B)$ in Sec. II. In Sec. III we prove the analytical form of the upper and lower bounds of coherent ergotropy. In Sec. IV we derive the condition under which diagonal entropy effects incoherent ergotropy. In Sec. V we further verify the propositions in the main text under higher-dimensional QBs. Finally, the propositions are exemplified via supplementary examples: a Jaynes-Cummings (JC) QB with a charger initially in a Fock state and an open Dicke QB model in Sec. VI.
\vspace{30pt}
\twocolumngrid
\maketitle
\section{THE RANDOM SAMPLING METHOD}\label{section1}
Here, we provide a detailed exposition of how we sample random states and Hamiltonians in our numerical simulations. The high-level procedure we follow is: (i) Generate a random $\rho_B$; (ii) Independently from $\rho_B$, generate a random $H_B$. The randomly generated state $\rho_B$ satisfies the following conditions: \cite{The_theory_of_open_quantum_systems},
\begin{equation} \label{cod_rho} 
    \rho^\dagger = \rho,\ \rho\geq 0,\ \mathrm{Tr}[\rho]=1.
\end{equation}
In addition, the random Hamiltonian $H$ is Hermitian.
\subsection{Random sampling states}\label{sec1.1}
\subsubsection{The Hilbert--Schmidt sampling method}\label{sec1.1.1}
In our analysis, random quantum states without any restrictions are sampled form the Hilbert--Schmidt measure. Although the Bures measure is often considered the ``most random" measure, it is skewed towards pure states. The Hilbert--Schmidt measure, by contrast, provides a ``flatter'' purity distribution and is a natural reduction of the Fubini-Study metric, thereby achieving good balance between randomness and a fair representation of all state ranks \citep{PhysRevLett.134.010408}. Numerically, a more efficient method for sampling random density matrix from the Hilbert--Schmidt measure is to randomly generate a $d \times d$ Ginibre matrix $G$ and construct
\begin{equation}
    \rho = \frac{GG^\dagger}{\operatorname{Tr}\left[GG^\dagger\right]}.
\end{equation}
As proven in Ref.~\citep{KarolZyczkowski_2001}, this procedure yields state $\rho$ indeed distributed according to the Hilbert--Schmidt measure. Here, we call the random sampling method based on the Hilbert-Schmidt measure as the Hilbert-Schmidt random sampling (HSRS) method. However, despite its attractive properties and widespread utilization, HSRS do not completely meet our requirements. Therefore, we propose two random sampling methods to cover the full range of diagonal entropy and purity, respectively.

\subsubsection{The Full-Entropy-Range Sampling}\label{sec1.1.2}
For incoherent ergotropy, we randomly generated the dephased state $\delta_B$ for simplicity without loss of generality, thereby converting the problem into random sampling of energy level populations.~The energy level populations of a dephased state sampled by HSRS, follow a Dirichlet distribution, $\operatorname{Dir} (2, 2, \dots, 2)$ \citep{Bengtsson_2017}.~However, the dephased states sampled are rarely found in regions of low diagonal entropy.

\begin{figure}
    \centering
    \includegraphics[width=0.485\textwidth]{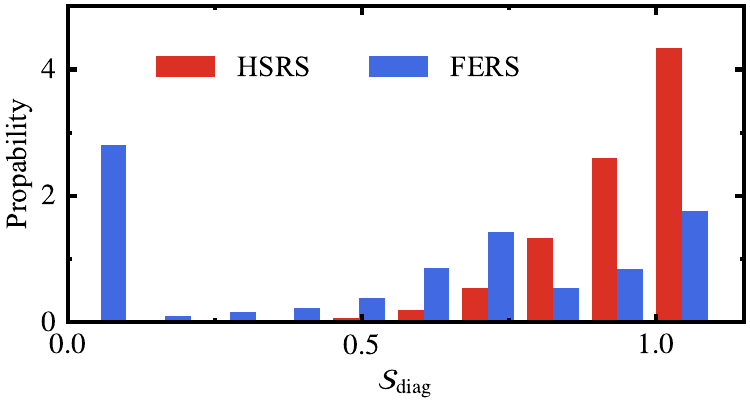}
    \caption{Distribution of the diagonal entropy $\mathcal{S}_{\text{diag}}$ for the two random sampling method for $d=3$. The red histogram originates from HSRS, while the blue one is generated by FERS algorithm with $N/N_{HS}=3/2$. Each histogram contains $10^7$ points.}
    \label{sfig1}
\end{figure}

To guarantee the validity of our analysis, we design a Full-Entropy-Range sampling (FERS) algorithm aiming to cover the full range of diagonal entropy.~This procedure consists of three steps: (i) drawing $N_{HS}$ dephased states by HSRS; (ii) randomly generating $N$ dephased states in the low diagonal entropy region; and (iii) controlling the sampling weight via the ratio $N/N_{HS}$. Since stronger localization often corresponds to lower diagonal entropy, we randomly sample dephased state drawing from the region of low diagonal entropy as the following procedure: (i) Select an integer $k$ uniformly from $\left\{1,2,\cdots,d-1\right\}$; (ii) Uniformly sample $k$ independent random numbers from the interval $\left[0,1\right)$ to form an array $\mathbf{x}=\left\{x_1,x_2,\cdots,x_k\right\}$; (iii) Append zeros to $\mathbf{x}$ until its length equals the dimension $d$; (iv) Shuffle the array $\mathbf{x}$ by the Fisher-Yates algorithm \citep{KnuthDonaldE}; and (v) Assign the elements of $\mathbf{x}$ sequentially to the diagonal of the density matrix (with zeros off-diagonal), i.e., $\rho_{nn}=x_n$ and normalize to unit trace. It is evident that the randomly generated density matrix satisfy physical constraint Eq. (\ref{cod_rho}). By repeating the process (i)-(v) $N$ times, we can obtain $N$ dephased states in the desired low-entropy region. Figure \ref{sfig1} demonstrates a comparison of the two random sampling methods for $d=3$. The FERS algorithm effectively increases the number of dephased states in the region of low diagonal entropy. Notably, the advantage of FERS becomes more pronounced in higher-dimensional cases.
\subsubsection{The Full-Purity-Range Sampling}\label{sec1.1.3}
The states generated from HSRS are strongly concentrated around purity $\mathcal{P} \approx 1/2$, whereas rarely appear in the regions near $\mathcal{P}=1$ (pure states) or $\mathcal{P}=1/d$ (the maximally mixed state), especially in higher-dimensional QBs \citep{Bengtsson_2017}. This distribution is inadequate for the purpose of to cover the entire purity range. Therefore, we propose the Full-Purity-Range Sampling (FPRS) algorithm. Further, the FPRS algorithm generates a random state with a target purity in two core steps: (i) Purity control, where the eigenvalue spectrum is engineered to achieve the desired purity; (ii) State randomization, where a density matrix is randomly generated from this spectrum via a unitary transformation sampled from the Haar measure \citep{Bengtsson_2017}.

\begin{figure}[htbp]
\centering
\includegraphics[width=0.485\textwidth]{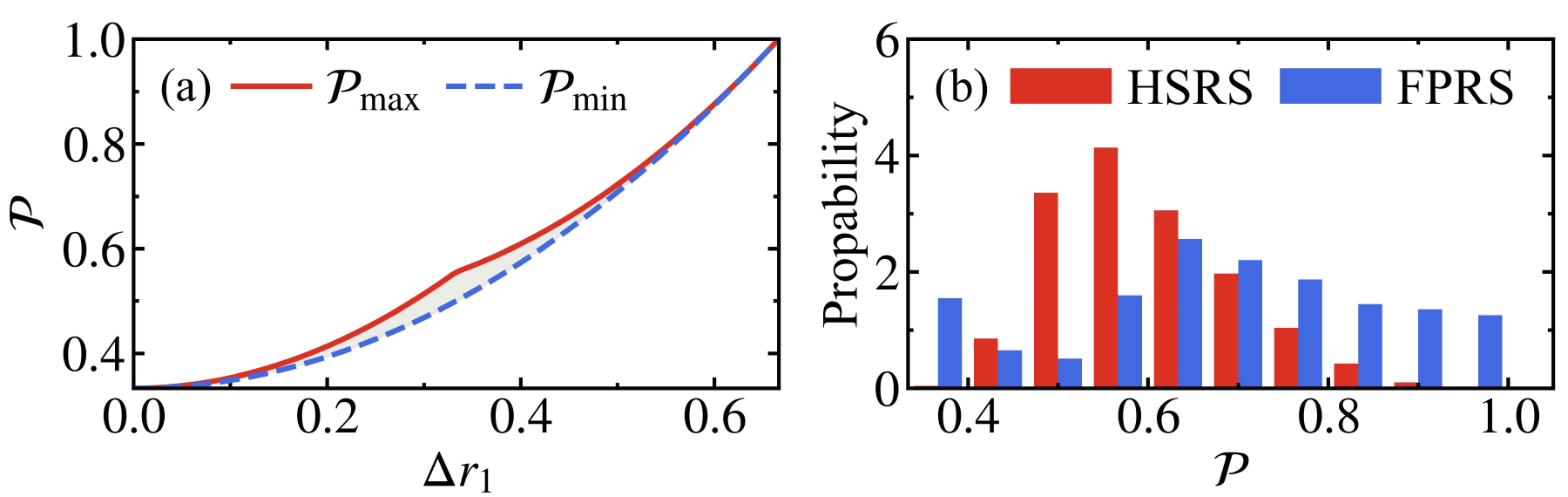}
\caption{(a) The dependence of purity on $\Delta r_1$ for $d=3$. The gray shaded region between the upper and lower bounds represents the accessible range of purity. (b) Distribution of the purity for two random sampling method for $d=3$. The red histogram originates from HSRS, while the blue one is generated by FPRS algorithm with $N_l:N_m:N_h=0.2:0.5:0.3$. Each histogram contains $10^7$ points.}
\label{sfig2}
\end{figure}

\emph{Purity control.}---The procedure for controlling the eigenvalues is as follows:~(i) Begin with the maximally mixed state, whose eigenvalues are $\mathbf{r}_0=\left\{1/d, 1/d, \cdots, 1/d\right\}$;~(ii) Introduce a random perturbation to the eigenvalues, $\Delta \mathbf{r}=\left\{\Delta r_1, -\Delta r_2, \cdots, -\Delta r_d\right\}$, with $\Delta r_n \geq 0$, thereby yielding a new eigenvalue spectrum $\mathbf{r}=\left\{1/d+\Delta r_1, 1/d-\Delta r_2, \cdots, 1/d-\Delta r_d\right\}$.~The  constraints of trace preservation and positive semidefiniteness require
\begin{equation}\label{cod_var}
\sum_{n=2}^d\Delta r_n=\Delta r_1,\quad 0\leq\Delta r_n \leq \frac{1}{d}\ (n\geq 2).
\end{equation}
(iii) The purity is controlled by tuning the parameter $\Delta r_1$. The dependence of the purity on the parameter $\Delta r_1$ is analyzed in detail below. For the spectrum $\mathbf{r}$, the purity is given by
\begin{equation}\label{pur_base}
\begin{aligned}
\mathcal{P} &= (1/d+\Delta r_1)^2+\sum_{n=2}^d(1/d-\Delta r_n)^2 \\
&= 1/d + \Delta r_1^2 + \sum_{n=2}^d \Delta r_n^2.
\end{aligned}
\end{equation}
By applying the Cauchy-Schwarz inequality \citep{Cauchy_Schwarz} and the constrains (\ref{cod_var}), we can obtain
\begin{equation}\label{cau_sch}
    \sum_{n=2}^d\Delta r_n^2 \geq \frac{1}{d-1}\left(\sum_{n=2}^d\Delta r_n\right)^2 = \frac{\Delta r_1^2}{d-1}.
\end{equation}
Substituting Eq. (\ref{cau_sch}) into the expression for purity [Eq. (\ref{pur_base})] yields lower bound of purity:
\begin{equation}\label{pur_low}
    \mathcal{P}_{\min} =  1/d + \Delta r_1^2 + \frac{\Delta r_1^2}{d-1}.
\end{equation}
The upper bound of purity for a given $\Delta r_1$ is obtained by maximizing $\sum_{n=2}^d\Delta r_n^2$ under the constrains in Eq. (\ref{cod_var}). This maximum occurs in the configuration where $m=\lfloor \Delta r_1 d\rfloor$ eigenvalues are assigned the maximal perturbation of $1/d$, a single eigenvalue is perturbed by $\Delta r_1-m/d$, and the remaining $d-m-2$ eigenvalues are unperturbed. The corresponding upper bound on purity is
\begin{equation}\label{pur_max}
    \mathcal{P}_{\max} = \frac{1}{d} + \Delta r_1^2+\frac{m}{d^2}+\left(\Delta r_1-\frac{m}{d}\right)^2.
\end{equation}
Fortunately, as illustrated in Fig. \ref{sfig2}(a), both the upper and lower bound of purity is monotonically increasing functions of  $\Delta r_1$, and the region between them is narrow. The feature allows us to approximately delineate the low-, medium-, and high-purity regions based solely on the lower bound of purity. Given that the purity ranges from $1/d$ to $1$, we divide this interval uniformly into three equal parts, corresponding to the low-purity region $[1/d,(d+2)/(3d)]$, medium-purity region $[(d+2)/(3d),(2d+1)/(3d)]$, and high-purity region $[(2d+1)/(3d),1]$, respectively. The corresponding ranges of the parameter $\Delta r_1$ for each purity region are given by
\begin{equation}
    \begin{cases}
        \begin{aligned}
            &\text{(i) low-purity region:}\quad \Delta r_1\in\left[0,\frac{d-1}{\sqrt{3}d}\right],\\
            &\text{(ii) medium-purity region:} \quad  \Delta r_1\in\left[\frac{(d-1)}{\sqrt{3}d},\frac{\sqrt{2}(d-1)}{\sqrt{3}d}\right],\\
            &\text{(iii) high-purity region:} \quad \Delta r_1\in\left[\frac{\sqrt{2}(d-1)}{\sqrt{3}d},\frac{d-1}{d}\right].
        \end{aligned}
    \end{cases}
\end{equation}
For any given $\Delta r_1$, other eigenvalue perturbations $\Delta r_n$ are randomly generated under the constraints given by Eq. (\ref{cod_var}). Therefore, by adjusting the value of $\Delta r_1$, we can effectively generate eigenvalue spectra within the target purity region.

\emph{State randomization}---A random density matrix with a given eigenvalue spectrum $\mathbf{r}=\{r_1,r_2,\dots,r_d\}$ is generated in a unitarily invariant, and maximally unbiased manner. Specifically, let $\Lambda = \operatorname{diag}(r_1,r_2,\dots,r_d)$ denotes the diagonal matrix formed by the prescribed eigenvalues. A unitary matrix $U$ is sampled from the unitary group according to the Haar measure, and the random density matrix is constructed as
\begin{equation}
\rho = U\Lambda U^\dagger.
\end{equation}
This method induces a unitarily invariant probability measure on the set of density matrix with the fixed eigenvalues spectrum.

We now perform random sampling of states in each target purity region (low, medium, and high) with $N_l$, $N_m$, and $N_h$ samples, respectively. For comparison, an ensemble of size $N$ ($N=N_l+N_m+N_h$) random states is also generated via the HSRS algorithm. Figure \ref{sfig2}(b) compares the purity distributions of the states generated by the FPRS and HSRS methods. The states generated via the HSRS method are strongly concentrated around a purity of $\mathcal{P} \approx 0.5$, with scarce representation in the regions near $\mathcal{P}=1$ or $\mathcal{P}=1/d$. In contrast, our FPRS method yields a distribution that spans the entire purity range. Furthermore, the relative sampling weight across different purity regions can be flexibly tuned by adjusting the ratios among $N_l$, $N_m$, and $N_h$.

For comparison with the original sampling method, i.e., the widely used HSRS method, here we term the FERS and FPRS weighted random sampling methods.

\subsection{Random sampling Hamiltonian}\label{sec1.2}
To ensure generality and fairness in Hamiltonian sampling, we adopt the Gaussian unitary ensemble (GUE), which provides the unique unbiased distribution of Hermitian matrices with statistically independent parameters and unitary invariance \citep{PhysRevLett.134.010408}. Following the standard methods, we sample matrices from the GUE as defined in \citep{Edelman_Rao_2005}:
\begin{equation}
    H = \frac{1}{2}\left(G+G^\dagger\right).
\end{equation}
In our analysis, we normalize the Hamiltonians to ensure comparability across different random instances. Furthermore, in incoherent ergotropy, although we impose an additional constraint of equally spaced energy levels, the conclusions hold for arbitrary Hamiltonians.

\section{The Relation of $E(\tilde{\delta}_B)$ and $PR$}\label{section2}

For the dephased state $\delta_B$, its corresponding passive state can be given by $\tilde{\delta}_B=\sum_{n=1}^d\tilde{p}_n|\varepsilon_n\rangle\langle \varepsilon_n|$, with populations ordered non-increasingly $\tilde{p}_n\geq \tilde{p}_{n+1}$. Therefore, stronger localization implies a larger population in the lower energy levels, generally resulting in a lower locked energy $E(\tilde{\delta}_B)$. Given that the participation ratio characterizes localization, Fig. \ref{sfig2} shows the dependence of $E(\tilde{\delta}_B)$ on it for two- and three-level QBs. The results clearly indicate that the locked energy generally increases with the participation ratio; this relation is strictly positive for two-level QBs.
\begin{figure}[htbp]
\centering
\includegraphics[width=0.485\textwidth]{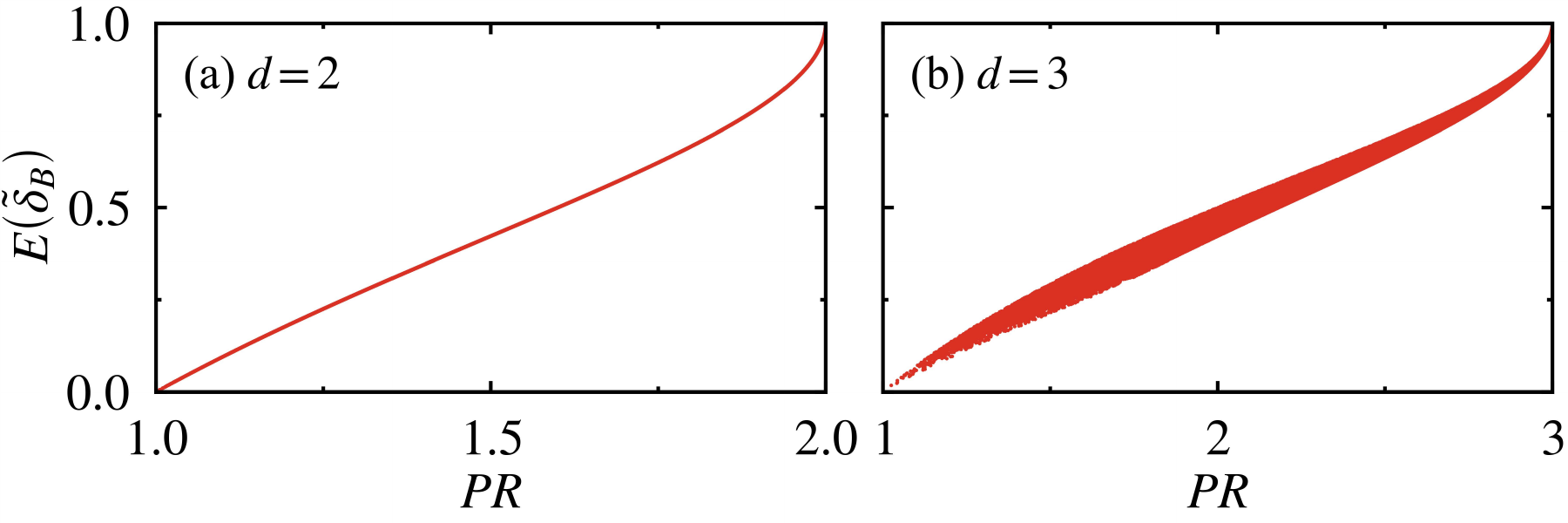}
\caption{(a),(b) The locked energy $E(\tilde{\delta}_B)$ of dephased state as a function with the participation ratio $PR$ for $d=2$ and $d=3$, respectively. We consider Hamiltonians with equally spaced and normalized energy levels. Each panel contains $10^7$ points.}
\label{sfig3}
\end{figure}
\section{The bounds on coherent ergotropy}\label{section3}
As shown by the Eq.~(10) in the main text, the coherent ergotropy satisfies the inequality
\begin{equation}
\mathcal{E}_c\left(\rho_B^{\text{pur}}\right)\leq \mathcal{E}_c\left(\rho_B\right)\leq \mathcal{E}_c\left(\rho_B^{\text{deloc}}\right),
\end{equation}
where the coherece for the pure and delocalized states are respectively given by $\mathcal{C}\left(\rho_B^{\text{pur}}\right)=-\sum_{n=1}^dp_n\log p_n=-\sum_{n=1}^d\tilde{p}_n\log \tilde{p}_n$ and $\mathcal{C}\left(\rho_B^{\text{deloc}}\right)=\log d+\sum_{n=1}^d r_n\log r_n$. To elucidate the relation between coherent ergotropy and coherence, we now rewrite the coherence of the pure state as
\begin{equation}\label{colow}
\begin{aligned}
    \mathcal{C}\left(\rho_B^{\text{pur}}\right)&=-\sum_{n=1}^d\chi_ne^{-\alpha \varepsilon_n}\left(\log\chi_n-\alpha \varepsilon_n\right)\\
    &=\alpha\sum_{n=1}^d\tilde{p}_n\varepsilon_n-\sum_{n=1}^d\tilde{p}_n\log \chi_n,\\
\end{aligned}
\end{equation}
where $\tilde{p}_n=\chi_n e^{-\alpha \varepsilon_n}$, with $\alpha$ and $\chi_n$ being auxiliary parameters. Considering the coherent ergotropy of pure state $\mathcal{E}_c\left(\rho_B^{\text{pur}}\right)=\sum_{n=1}^d\tilde{p}_n\varepsilon_n-\varepsilon_1$, we can obtain the low bound of coherent ergotropy
\begin{equation}\label{coerlow}
    \mathcal{E}_c\left(\rho_B^{\text{pur}}\right)=\frac{1}{\alpha}\left[\mathcal{C}\left(\rho_B^{\text{pur}}\right)+\sum_{n=1}^d \tilde{p}_n\log\chi_n -\alpha \varepsilon_1\right].
\end{equation}
Similarly, the coherence of the completely delocalized state can be rewritten as:
\begin{equation}
    \begin{aligned}
        \mathcal{C}\left(\rho_B^{\text{deloc}}\right)&=\log d+\sum_{n=1}^{d}r_n\log r_n\\
        &=\log d+\sum_{n=1}^dq_n e^{-\alpha \varepsilon_n}\left(\log q_n-\alpha \varepsilon_n\right)\\
        &=\log d+\sum_{n=1}^d r_n\log q_n-\alpha\sum_{n=1}^d r_n\varepsilon_n,
    \end{aligned}
\end{equation}
where $r_n=q_ne^{-\alpha \varepsilon_n}$ with $q_n$ being an auxiliary parameter. Furthermore, considering the coherent ergotropy of completely delocalized state $\mathcal{E}_c\left(\rho_B^{\text{deloc}}\right)=\sum_{n=1}^d\left(1/d-r_n\right)\varepsilon_n$, the upper bound of coherent ergotropy can be given by
\begin{equation}\label{coerup}
\begin{aligned}
    \mathcal{E}_c\left(\rho_B^{\text{deloc}}\right)&=\frac{1}{\alpha}\Big[\mathcal{C}\left(\rho_B^{\text{deloc}}\right)+\alpha\mathrm{Tr}\left(H_B\rho_B^{\text{deloc}}\right)\\
    &-\log d-\sum_{n=1}r_n\log q_n\Big].
\end{aligned}
\end{equation}

\section{Condition of Diagonal Entropy Affecting Incoherent Ergotropy}\label{section4}
In this section, we analyze how the incoherent ergotropy varies with the diagonal entropy in different ordering of level populations for three-level QBs whose Hamiltonian has equally spaced energy levels and is normalized. We consider the population distribution of the state $\rho_B$: $p_n = \langle \varepsilon_n | \rho_B | \varepsilon_n \rangle$, which necessarily satisfies a specific ordering: $p_i \leq p_j \leq p_k$ $(i,j,k=1,2,3)$. After a short time of evolution, the state of QBs becomes $\rho_B'$, with a slight variation in population: $p_n'=\langle \varepsilon_n|\rho_B'|\varepsilon_n\rangle$, where $\Delta p_n=p_n'-p_n$, and we assume that this small variation $\Delta p_n$ dose not alter the original ordering. Due to $\sum_{n=1}^3 p'_n=1$, we obtain $\Delta p_k=-\Delta p_i-\Delta p_j$. We now focus on the variation in the diagonal entropy with the population:
\begin{equation}
\begin{aligned}\label{DS} 
\mathbf{grad}\ &{}\mathcal{S}_{\text{diag}}\left(p_i,p_j,p_k\right) \cdot \left(\Delta p_i,\Delta p_j,\Delta p_k\right)\\
&=\Delta p_i\log\frac{p_k}{p_i}+\Delta p_j\log\frac{p_k}{p_j},
\end{aligned}
\end{equation}
where $\mathcal{S}_{\text{diag}}\left(p_i,p_j,p_k\right)=\mathcal{S}_{\text{diag}}\left(\rho_B\right)$, $\mathbf{grad}\ \mathcal{S}_{\text{diag}}$ denotes the gradient of the diagonal entropy. Further, we can obtain
{\setlength{\jot}{6pt}
\begin{equation}\label{conS} 
\left\{
\begin{array}{rl}
\mathcal{S}_{\text{diag}}\left(\rho_B'\right)>\mathcal{S}_{\text{diag}}\left(\rho_B\right) & \text{if }\ \Delta p_i >-\Delta p_j\frac{\log p_k/p_j}{\log p_k/p_i},\\
\mathcal{S}_{\text{diag}}\left(\rho_B'\right)=\mathcal{S}_{\text{diag}}\left(\rho_B\right) & \text{if }\ \Delta p_i =-\Delta p_j\frac{\log p_k/p_j}{\log p_k/p_i},\\
\mathcal{S}_{\text{diag}}\left(\rho_B'\right)<\mathcal{S}_{\text{diag}}\left(\rho_B\right) & \text{if }\ \Delta p_i <-\Delta p_j\frac{\log p_k/p_j}{\log p_k/p_i}.
\end{array}
\right.
\end{equation}}

We now turn to the incoherent ergotropy. Since the incoherent ergotropy depends on the ordering of level populations, we first investigate  incoherent ergotropy in the case of global population inversion (the $\mathrm{\Rmnum{3}}$ region of Fig. 1 in the main text with $p_1\leq p_2\leq p_3$ and $p_3>p_1$). The incoherent ergotropy is given by $\mathcal{E}_i=2\left(p_3-p_2\right)$. The variation of incoherent ergotropy is $\mathcal{E}_{i}\left(\rho_B'\right)-\mathcal{E}_{i}\left(\rho_B\right)=2\left(\Delta p_3-\Delta p_2\right)$, from which we obtain
\begin{equation}\label{conEi} 
\left\{
\begin{array}{rl}
\mathcal{E}_i\left(\rho_B'\right)>\mathcal{E}_i\left(\rho_B\right)&\text{if }\Delta p_1<-\frac{1}{2}\Delta p_2,\\
\mathcal{E}_i\left(\rho_B'\right)=\mathcal{E}_i\left(\rho_B\right)&\text{if }\Delta p_1=-\frac{1}{2}\Delta p_2,\\
\mathcal{E}_i\left(\rho_B'\right)<\mathcal{E}_i\left(\rho_B\right)&\text{if }\Delta p_1>-\frac{1}{2}\Delta p_2.
\end{array}
\right.
\end{equation}
To elucidate the effect of the diagonal entropy on the incoherent ergotropy, we refer to table~\ref{table1}.
\begin{table}[hbtp]
\renewcommand{\arraystretch}{1.5}
\setlength{\abovecaptionskip}{0pt}   
\setlength{\belowcaptionskip}{6pt}   
\centering
\caption{\label{table1} The conditions of the diagonal entropy affecting the incoherent ergotropy in the case of the global population inversion ($p_1\leq p_2\leq p_3$ and $p_3>p_1$). The $\Delta \mathcal{S}_{\text{diag}}$ and $\Delta \mathcal{E}_i$ denote $\mathcal{S}_{\text{diag}}\left(\rho_B'\right)-\mathcal{S}_{\text{diag}}\left(\rho_B\right)$ and $\mathcal{E}_{i}\left(\rho_B'\right)-\mathcal{E}_{i}\left(\rho_B\right)$, respectively. The symbols $+$,$0$ and $-$ represent $\Delta \mathcal{S}_{\text{diag}}$ ($\Delta \mathcal{E}_{i}$) greater than, equal to and lower than zero, respectively.
}
\begin{tabularx}{\linewidth}{@{} >{\centering\arraybackslash}X |>{\centering\arraybackslash}p{0.9cm}| >{\centering\arraybackslash}p{0.6cm}| >{\centering\arraybackslash}p{1.4cm}@{}}
    \hline\hline
    Condition & $\Delta\mathcal{S}_{\text{diag}}$ & $\Delta\mathcal{E}_i$ & Effect\\ \hline
    $-\Delta p_2\log\frac{p_3}{p_2}/\log\frac{p_3}{p_1}<\Delta p_1<-\frac{1}{2}\Delta p_2$& $+$& $+$&\\
    $-\frac{1}{2}\Delta p_2<\Delta p_1<-\Delta p_2\log\frac{p_3}{p_2}/\log\frac{p_3}{p_1}$& $-$ & $-$&\multirow{-2}{*}{enhance}\\
    \hline
     $\Delta p_1>\max\left\{-\frac{1}{2}\Delta p_2,-\Delta p_2\log\frac{p_3}{p_2}/\log\frac{p_3}{p_1}\right\}$& $+$& $-$&\\
    $\Delta p_1<\min\left\{-\frac{1}{2}\Delta p_2,-\Delta p_2\log\frac{p_3}{p_2}/\log\frac{p_3}{p_1}\right\}$& $-$ & $+$&\multirow{-2}{*}{suppress}\\
    \hline
    $\Delta p_1=-\frac{1}{2}\Delta p_2>-\Delta p_2\log\frac{p_3}{p_2}/\log\frac{p_3}{p_1}$& $+$&$0$& \\
    $\Delta p_1=-\frac{1}{2}\Delta p_2<-\Delta p_2\log\frac{p_3}{p_2}/\log\frac{p_3}{p_1}$& $-$ & $0$&\multirow{-2}{*}{maintain}\\ \hline
    $\Delta p_1=-\Delta p_2\log\frac{p_3}{p_2}/\log\frac{p_3}{p_1}<-\frac{1}{2}\Delta p_2$& $0$&$+$&\\
    $\Delta p_1=-\Delta p_2\log\frac{p_3}{p_2}/\log\frac{p_3}{p_1}>-\frac{1}{2}\Delta p_2$& $0$&$-$&\multirow{-2}{*}{other}\\
 \hline\hline
\end{tabularx}
\end{table}

The first and second lines demonstrate the enhancement of incoherent ergotropy by diagonal entropy. The fifth and sixth lines describe variation in diagonal entropy with constancy in incoherent ergotropy, while the seventh and eighth lines depict constancy in diagonal entropy with variation in incoherent ergotropy. To achieve these three scenarios, the populations must satisfy specific relationships. For instance, when $\Delta \mathcal{S}_{\text{diag}} > 0$ and $\Delta \mathcal{E}_i > 0$, the condition
\begin{equation*}
\left\{
\begin{array}{cc}
    2\log \frac{p_3}{p_2}>\log \frac{p_3}{p_1}& \text{if }\Delta p_2>0,\\
    2\log \frac{p_3}{p_2}<\log \frac{p_3}{p_1}& \text{if }\Delta p_2<0,\\
\end{array}
\right.
\end{equation*}
must be met. The third and fourth lines show that diagonal entropy can suppress incoherent ergotropy. This suppression is particularly facilitated under global population inversion because it requires no additional constraints on the populations. These analyses provide a clear explanation for the various changes in incoherent ergotropy depicted in Fig. 2(d) of the main text.

In the case of local population inversion, we consider the ordering $p_1 \geq p_3 > p_2$ [the $\mathrm{\Rmnum{2}}_{1}$ region of Fig. 1(c) in main text], where the incoherent ergotropy is given by $\mathcal{E}_i\left(\rho_B\right) = p_3 - p_2$. The variation in incoherent ergotropy is then $\mathcal{E}_i\left(\rho_B'\right) - \mathcal{E}_i\left(\rho_B\right) = \Delta p_3 - \Delta p_2$, yielding the following relation:
\begin{equation}
\left\{
\begin{array}{cc}
\mathcal{E}_i\left(\rho_B'\right) > \mathcal{E}_i\left(\rho_B\right) & \text{if} \quad \Delta p_3 > \Delta p_2,\\
\mathcal{E}_i\left(\rho_B'\right) = \mathcal{E}_i\left(\rho_B\right) & \text{if} \quad \Delta p_3 = \Delta p_2,\\
\mathcal{E}_i\left(\rho_B'\right) < \mathcal{E}_i\left(\rho_B\right) & \text{if} \quad \Delta p_3 < \Delta p_2.\\
\end{array}
\right.
\end{equation}
Similarly, to explore the role of diagonal entropy in incoherent ergotropy, we refer to table \ref{table2}.

\begin{table}[hbtp]
\renewcommand{\arraystretch}{1.5}
\setlength{\abovecaptionskip}{0pt}   
\setlength{\belowcaptionskip}{6pt}   
\centering
\caption{\label{table2} The conditions of the diagonal entropy affecting the incoherent ergotropy in a ordering ($p_1\geq p_3>p_2$). The meaning of the other symbols are consistent with those defined in table \ref{table1}.}
\begin{tabularx}{\linewidth}{@{} >{\centering\arraybackslash}X |>{\centering\arraybackslash}p{0.9cm}| >{\centering\arraybackslash}p{0.9cm}| >{\centering\arraybackslash}p{1.5cm}@{}}
    \hline\hline
    Condition & $\Delta\mathcal{S}_{\text{diag}}$ & $\Delta\mathcal{E}_i$ & Effect\\ \hline
    $-\Delta p_3\log\frac{p_1}{p_3}/\log\frac{p_1}{p_2}<\Delta p_2<\Delta p_3$&$+$&$+$&\\
    $\Delta p_3<\Delta p_2<-\Delta p_3\log\frac{p_1}{p_3}/\log\frac{p_1}{p_2}$&$-$&$-$&\multirow{-2}{*}{enhance}\\ \hline
    $\Delta p_2>\max\left\{\Delta p_3,-\Delta p_3\log\frac{p_1}{p_3}/\log\frac{p_1}{p_2}\right\}$&$+$&$-$&\\
    $\Delta p_2<\min\left\{\Delta p_3,-\Delta p_3\log\frac{p_1}{p_3}/\log\frac{p_1}{p_2}\right\}$&$-$&$+$&\multirow{-2}{*}{suppress}\\ \hline
    $\Delta p_2=\Delta p_3>-\Delta p_3\log\frac{p_1}{p_3}/\log\frac{p_1}{p_2}$&$+$&$0$&\\
    $\Delta p_2=\Delta p_3<-\Delta p_3\log\frac{p_1}{p_3}/\log\frac{p_1}{p_2}$&$+$&$0$&\multirow{-2}{*}{maintain}\\ \hline
    $\Delta P_2=-\Delta p_3\log\frac{p_1}{p_3}/\log\frac{p_1}{p_2}<\Delta p_3$&$0$&$+$&\\
    $\Delta p_2=-\Delta p_3\log\frac{p_1}{p_3}/\log\frac{p_1}{p_2}>\Delta p_3$&$0$&$-$&\multirow{-2}{*}{other}\\
 \hline\hline
\end{tabularx}
\end{table}

In contrast to the cases of global population inversion, realizing any of these effects does not require additional constraints on the populations. Consequently, compared to the case of global population inversion, the incoherent ergotropy is more readily enhanced or maintained by the diagonal entropy. For other population ordering, applying the above analytical procedure reveals that incoherent ergotropy can be enhanced, maintained or suppressed by diagonal entropy.

\section{The propositions in higher dimension QBs}\label{section5}
In this section, we demonstrate the universality of the propositions in the main text in higher dimension QBs. Concretely, we investigate the coherent and incoherent ergotropy, charging efficiency, and locked energy for four- and five-level QBs.
\subsection{The coherent and incoherent ergotropy in higher dimension QBs}\label{sec5.1}
As mentioned in the main text, our goal is to pursue a general principle for the ergotropy of a model-independent QBs. Therefore, in this section, we illustrate proposition 2 in the higher-dimensional QBs. Figure \ref{sfig3} demonstrates the relation between quantum resource and both the coherent and incoherent ergotropy for four- and five-level QBs. For higher dimension QBs, the coherent ergotropy still increases with both the coherence and participation ratio in Figs. \ref{sfig3}(a) and \ref{sfig3}(b). The coherent ergotropy is bounded as $\mathcal{E}(\rho^{\text{pur}})\leq \mathcal{E}(\rho)\leq \mathcal{E}(\rho^{\text{deloc}})$ when the coherence is equal, i.e., $\mathcal{C}(\rho^{\text{pur}})=\mathcal{C}(\rho)=\mathcal{C}(\rho^{\text{deloc}})$. Figures \ref{sfig3}(c) and \ref{sfig3}(d) depict the dependence of the incoherent ergotropy on both the diagonal entropy and the participation ratio for four- and five-level QBs, respectively. The incoherent ergotropy can be enhanced, maintained, or suppressed under different conditions, as analysed by the method in Sec.~IV. When the diagonal entropy (participation ratio) reaches its minimum, the density matrix localizes in any eigenstate $\mid \varepsilon_n\rangle$, i.e., $\mathcal{E}_i=\varepsilon_n-\varepsilon_1$. In the opposite limit of maximum diagonal entropy (participation ratio), the populations are equally distributed among all energy levels, resulting in $\mathcal{E}_i=0$.
\begin{figure}[htbp]
\centering
\includegraphics[width=0.485\textwidth]{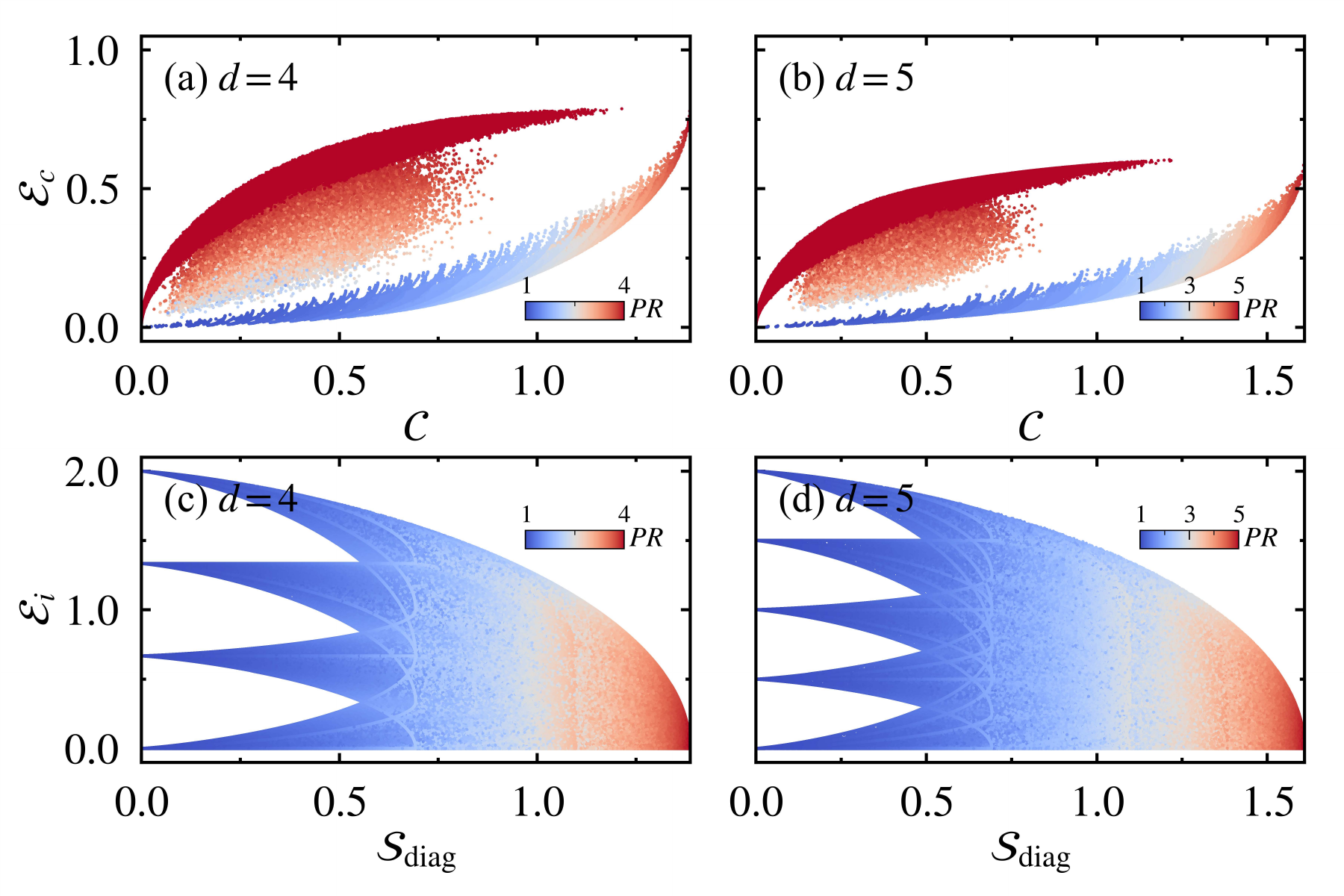}
\caption{(a),(b) The coherent ergotropy $\mathcal{E}_c$ as a function of the coherence $\mathcal{C}$ and participation ratio $PR$ for randomly sampled state $\rho_B$, Hamiltonian $H_B$. Here, the Hamiltonian $H_B$ is normalized. (c),(d) The incoherent ergotropy as a function of the diagonal entropy and participation ratio for randomly sampled states $\rho_B$, Hamiltonian $H_B$. We consider Hamiltonians with equally spaced and normalized energy levels. Each panels contains $10^7$ points. Undeniably, as the QBs dimension increases, the HSRS method gradually becomes inadequate for sampling in the high-coherence regions.}
\label{sfig4}
\end{figure}
\subsection{The charging efficiency and locked energy in higher dimension QBs}\label{sec5.2}
\begin{figure}[htbp]
\centering
\includegraphics[width=0.485\textwidth]{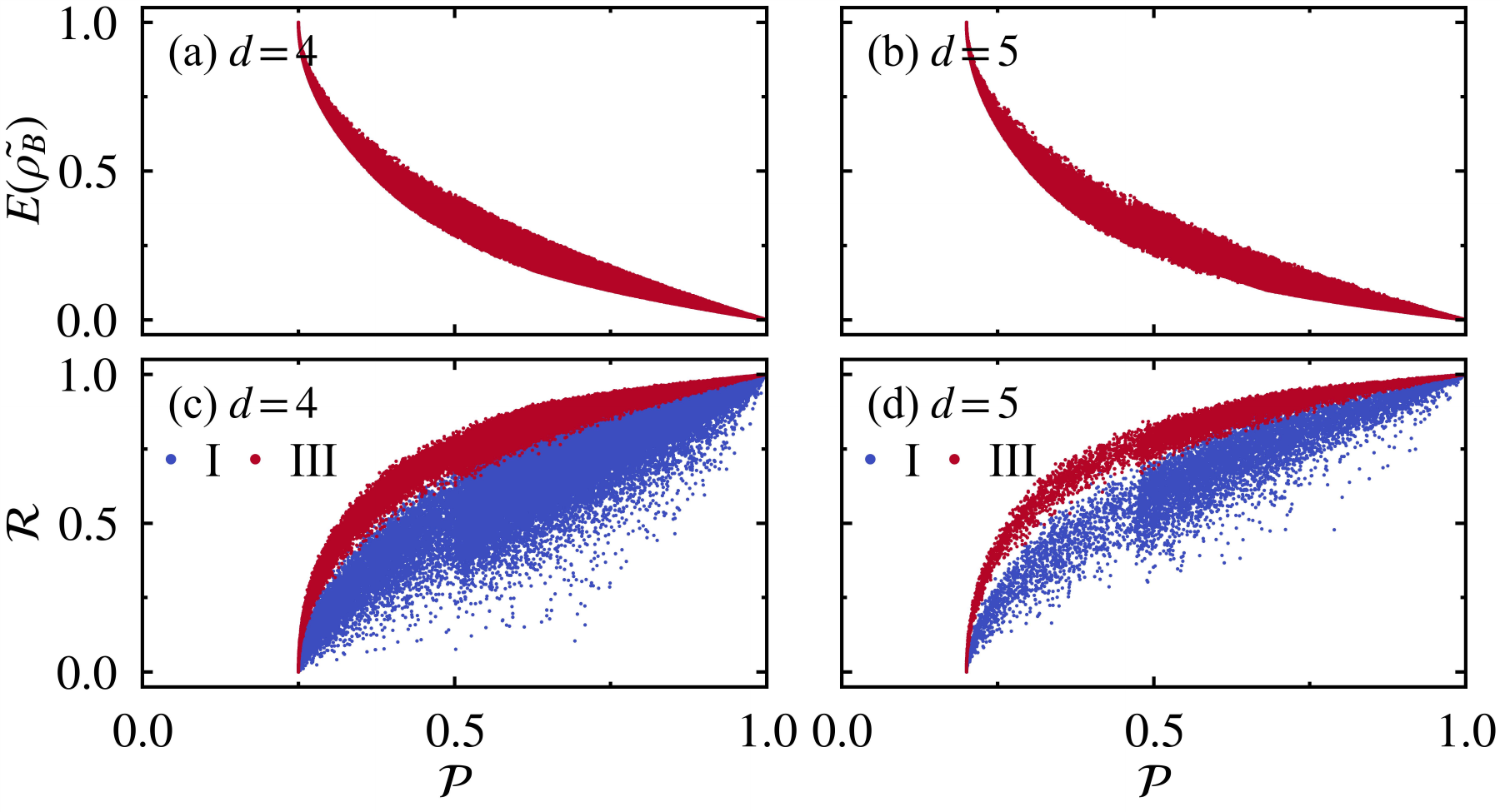}
\caption{The dependence of both (a),(b) the locked energy $E(\tilde{\rho})$ and (c),(d) the charging efficiency $\mathcal{R}$ on purity $\mathcal{P}$ for four- and five-level QBs. In (d), the symbols $\mathrm{\Rmnum{1}}$ and $\mathrm{\Rmnum{3}}$ denote the same configuration as in Fig. 1(c) in the main text. Each panel contains $10^5$ randomly sampled states.}
\label{sfig5}
\end{figure}

Here, we employ FPRS algorithm to demonstrate that the purity generally enhances charging efficiency and suppresses the locked energy in four- and five-level QBs. For a passive state $\tilde{\rho}_B$, the participation ratio satisfies $\mathcal{P}(\tilde{\rho}_B) = 1/PR(\tilde{\rho}_B)$. A higher purity $\mathcal{P}(\tilde{\rho}_B)$ thus implies a stronger localization of $\tilde{\rho}_B$ in the lower energy levels, which generally suppresses its locked energy $E(\tilde{\rho}_B)$, as shown in Figs. \ref{sfig5}(a) and \ref{sfig5}(b). Since the charging efficiency is defined as $\mathcal{R} = 1 - E(\tilde{\rho}_B)/E(\rho_B)$, this suppression of locked energy may enhance $\mathcal{R}$. We demonstrate the dependence of the charging efficiency on purity for four- and five-level QBs under the no population inversion and the global population inversion in Figs. \ref{sfig5}(c) and \ref{sfig5}(d), respectively. Although there is no ``deterministic'' relation between $\mathcal{R}$ and $\mathcal{P}$, a higher purity is generally necessary to achieve higher charging efficiency. Moreover, the enhancement of purity on charging efficiency becomes increasingly pronounced with the continuous population transfer from lower to higher energy levels, which brings the population distribution of QBs closer to the global population inversion.

\section{Exemplification of the Propositions}\label{section6}
In this section, we further present examples to validate the propositions in the main text. Specifically, we examine the JC QBs with a charger initialized in a Fock state and the open Dicke QBs. We present the dynamics of the ergotropy, as well as the dependence of its coherent and incoherent components on quantum resources. Furthermore, we demonstrate the relation of both the charging efficiency and the locked energy with purity.
\subsection{The JC QBs with a charger initially in a Fock state}\label{sec6.1}
The JC battery Hamiltonian is given by
\begin{equation}\label{TCH}
\begin{aligned}
H=H_B+H_C+\lambda\left(t\right)\mathcal{V},\\
H_B=\frac{\omega_b}{2}\sigma_z,\quad H_C=\omega_c a^\dagger a\\
\mathcal{V}=g\left(\sigma_+a+\sigma_-a^\dagger\right),
\end{aligned}
\end{equation}
where $H_B,H_C$ and $\mathcal{V}$ represent the Hamiltonian of the battery, charger and their interaction, respectively. Here, $ \sigma_j^\alpha/2\ (\alpha=x,y,z)$ represent the Pauli operators and $\sigma_\pm=\sigma_x\pm i\sigma_y$ represent the ladder operators. $\lambda(t)$ is the step function, $\lambda(t)=1, t \in [0,T]$ and $\lambda(t)=0, \text{otherwise}$. We consider the resonant situation, specifically $\omega_b=\omega_c=\omega$.
The initial state of the total system reads $|\Psi(0)\rangle=|g\rangle_b\otimes\sum_n\alpha_n|n\rangle_c$, where $|n\rangle_c$ is the Fock state with $n$ photons and $\alpha_n$ is probability amplitudes. The wave function can be obtained at any time $t$,
\begin{equation}\label{psit}
\begin{aligned}
|\psi\left(t\right)\rangle&=e^{-iHt}|\psi\left(0\right)\rangle\\
&=\sum_n\alpha_n\Big[-i\sin\left(\frac{\Omega_nt}{2}\right)|e,n\rangle\\&
+\cos\left(\frac{\Omega_nt}{2}\right)|g,n+1\rangle\Big],
\end{aligned}
\end{equation}
where the $\Omega_n=2g\sqrt{n+1}$ is Rabi frequency. To derive the explicit form of the ergotropy for our system, diagonalization of the density matrix is required. Its eigenvalues are
\begin{equation}\label{eigvalues}
r_s\left(t\right)=\frac{1-\left(-1\right)^{s}\sqrt{1-4\det\rho_B\left(t\right)}}{2},
\end{equation}
where $s=1,2$ with $r_1\leq r_2$ for any time $t$. Therefore, we can obtain the ergotropy of QB
\begin{equation}\label{ergotropy}
\mathcal{E}(t)=-\frac{\hbar\omega}{2}\sum_n\vert \alpha_n\vert^2\cos\left(\Omega_nt\right)+\frac{\hbar\omega}{2}\sqrt{1-4\det\rho_B}.
\end{equation}
We consider the charger initialized in a Fock state, described by the probability amplitudes $\alpha_n^{(F)}=\delta_{n,N_c}$ with an average number $N_c$ of photons. The corresponding ergotropy from Eq. (\ref{ergotropy}) is
\begin{equation}\label{erfc}
\mathcal{E}\left(t\right)=
\begin{cases}
0 & \text{if }\quad t>3T_{N_c}/4 \text{ or } t<T_{N_c}/4,\\
-\hbar\omega\cos\left(\Omega_{N_c}t\right)& \text{if }\quad T_{N_c}/4\leq t\leq 3T_{N_c}/4.
\end{cases}
\end{equation}
\begin{figure}[htbp]
\centering
\includegraphics[width=0.485\textwidth]{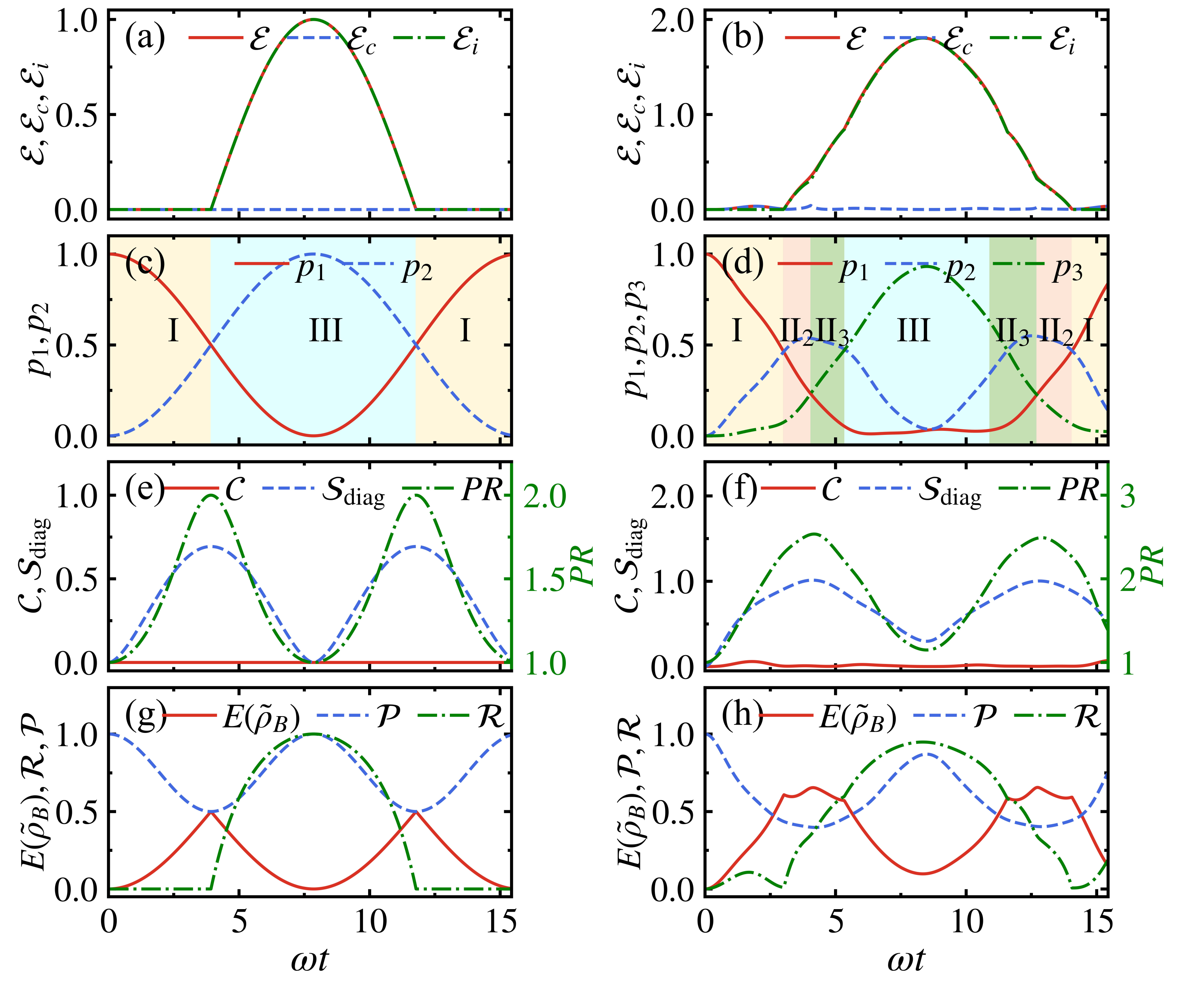}
\caption{Time evolution of (a),(b) ergotropy, coherent ergotropy, incoherent ergotropy, (c),(d) energy level populations, (e),(f) locked energy, purity, charging efficiency. (a), (c), (e) and (g) correspond to the cases of JC QBs with a charger initially in $|N_c\rangle_c$, and (b), (d), (f) and (h) correspond to the cases of the open Dicke QBs with two two-level systems ($N_B=2$), which is analogous to the three-level QBs. The symbols $\mathrm{\Rmnum{1}}$, $\mathrm{\Rmnum{2}}_{1}$, $\mathrm{\Rmnum{2}}_{2}$, $\mathrm{\Rmnum{2}}_{3}$, $\mathrm{\Rmnum{2}}_{4}$ and $\mathrm{\Rmnum{3}}$ denote the same configurations as in Fig. 1(c) of the main text. Other parameters are set to $\omega=1$, $N_c=4$, $g=0.1$ and $\kappa=0.5$.}
\label{sfig6}
\end{figure}
Here, this ergotropy $\mathcal{E}(t)$ is a periodic function of time with period $T_{N_c} = \pi/(g\sqrt{N_c+1})$. According to Eq. (\ref{psit}), the coherence of battery is always zero during the charging process in this cases, as illustrated in Fig. \ref{sfig6}(e). Therefore, the ergotropy coincides with incoherent ergotropy; see Fig. \ref{sfig6}(a). In the intervals $\left[0,T_{N_c}/4\right]$ or $\left[3T_{N_c}/4,T_{N_c}\right]$, the ergotropy vanishes due to the absence of both population inversion and coherence. In contrast, within $\left[T_{N_c}/4,3T_{N_c}/4\right]$, the ergotropy is greater than zero and increases with the population of the excited state, as shown in Figs. \ref{sfig6}(a) and \ref{sfig6}(c). At time $t=T_{N_c}/2$, the battery is fully charged and capable of extracting all of its stored energy. The incoherent ergotropy is negatively related to the diagonal entropy or participation ratio in cases of global population inversion [see Figs. \ref{sfig6}(a) and \ref{sfig6}(e)]. In addition, the purity $\mathcal{P}$ suppresses the locked energy $E\left(\tilde{\rho}\right)$ and enhances charging efficiency $\mathcal{R}$ [see Fig. \ref{sfig6}(g)].

\subsection{The open Dicke QBs}\label{sec6.2}
For the Dicke QB constituted by $N_B$ two-level systems (TLSs) coupled to a single-mode cavity, the overall system can be described by the following Hamiltonian:
\begin{equation}\label{dickeH}
    H=\omega_c a^\dagger a+\omega_b J_z+2\omega_cgJ_x\left(a^\dagger +a\right),
\end{equation}
where $J_\alpha = 1/2\sum_{i=1}^N\sigma_i^\alpha$ is the components of a collective spin operator in terms of the Pauli operator $\sigma_{i}^\alpha$. We consider the resonant case, i.e., $\omega_c=\omega_b=\omega$. For the open Dicke QBs, its dynamics can be described by the following master equation
\begin{equation}
\dot{\rho}(t)=\frac{1}{i\hbar}\left[H,\rho\left(t\right)\right]+\kappa\mathcal{D}_{a}\left[\rho\left(t\right)\right],
\end{equation}
where $\kappa$ is the damping rate and $\mathcal{D}_A[\bullet]$ represents the superoperator
\begin{equation}\label{superoper}
\mathcal{D}_A[\bullet]=A\bullet A^\dagger-\frac{1}{2}\{A^\dagger A,\bullet\}.
\end{equation}
The Dicke QB is initialized in the state $|\psi(0)\rangle = |0\rangle_b \otimes |N_c\rangle_c$. Here, $|0\rangle_b = |g,\dots,g\rangle_b$ denotes all TLSs in their ground state, and $|N_c\rangle_c$ is the Fock state of charger with $N_c$ photons.

Time evolution of ergotropy, coherent ergotropy, incoherent ergotropy are shown in Fig. \ref{sfig6}(b). Combining the energy level population [see Fig. \ref{sfig6}(d)], as described by Proposition $1$ in the main text, the ergotropy coincides with coherent ergotropy without the population inversion; otherwise, the ergotropy is determined by coherent and incoherent ergotropy with the population inversion. Further, the coherent ergotropy generally is enhanced by the coherence and participation ratio. The incoherent ergotropy may be enhanced or suppressed by diagonal entropy or participation ratio [see Figs. \ref{sfig6}(b) and \ref{sfig6}(f)]. Similarly, the purity generally enhances the charging efficiency and suppresses the locked energy [see Fig. \ref{sfig6}(h)].

\end{document}